\documentclass[10pt, journal]{IEEEtran}
\ifCLASSINFOpdf
\else
   \usepackage[dvips]{graphicx}
\fi
\usepackage{url}
\usepackage{cite}
\usepackage{color}
\usepackage{graphicx}
\usepackage{epstopdf}
\usepackage{changes}
\usepackage{amsmath}
\usepackage{subcaption}
\usepackage{float}
\usepackage{siunitx} 
\usepackage[english]{babel}
\usepackage[utf8]{inputenc}
\usepackage{algorithm} 
\usepackage{algpseudocode} 
\usepackage{bbding}
\usepackage{pifont}
\usepackage{wasysym}
\usepackage{amssymb}
\usepackage{multirow}
\newtheorem{remark}{Remark}
\newcommand{\xmark}{\ding{55}}

\begin{document}
\setlength{\abovedisplayskip}{4pt}
\setlength{\belowdisplayskip}{4pt}

\title{\huge Machine Learning-based Urban Canyon Path Loss Prediction using 28~GHz Manhattan Measurements}
\author{Ankit~Gupta,~\IEEEmembership{Student Member,~IEEE,}
        Jinfeng~Du,~\IEEEmembership{Member,~IEEE,}
        Dmitry~Chizhik,~\IEEEmembership{Fellow,~IEEE,}
        Reinaldo~A.~Valenzuela,~\IEEEmembership{Fellow,~IEEE,}
        and~Mathini~Sellathurai,~\IEEEmembership{Senior Member,~IEEE}
        \vspace*{-0.25cm}
\thanks{Ankit Gupta and Mathini Sellathurai are with the Engineering and physical science (EPS) department at Heriot-Watt University, Edinburgh, UK, (e-mail: \{ag104, m.sellathurai\}@hw.ac.uk).}
\thanks{Jinfeng Du, Dmitry Chizhik and Reinaldo A. Valenzuela are with Nokia Bell Laboratories, Murray Hill, NJ 07974, USA, (e-mail: \{jinfeng.du, dmitry.chizhik, reinaldo.valenzuela\}@nokia-bell-labs.com). }
\thanks{The authors from Heriot-Watt University are supported by the U.K. Engineering and Physical Sciences Research Council (MANGO) under Grant EP/P009670/1. A. Gupta is also supported by the Doctoral training program under the Grant EP/N509474/1-1963633.}}

\maketitle 

\begin{abstract} 
Large bandwidth at mm-wave is crucial for 5G and beyond but the high path loss (PL) requires highly accurate PL prediction for network planning and optimization. {Statistical models with slope-intercept fit fall short in capturing large variations seen in urban canyons, whereas ray-tracing, capable of characterizing site-specific features, faces challenges in describing foliage and street clutter and associated reflection/diffraction ray calculation.} Machine learning (ML) is promising but faces three key challenges in PL prediction: 1) insufficient measurement data; 2) lack of extrapolation to new streets; 3) overwhelmingly complex features/models. We propose an ML-based urban canyon PL prediction model based on extensive 28 GHz measurements from Manhattan where street clutters are modeled via a LiDAR point cloud dataset and buildings by a mesh-grid building dataset. We extract expert knowledge-driven street clutter features from the point cloud and aggressively compress 3D-building information using convolutional-autoencoder. Using a new street-by-street training and testing procedure to improve generalizability, the proposed model using both clutter and building features achieves a prediction error (RMSE) of $4.8 \pm 1.1$~dB compared to $10.6 \pm 4.4$~dB and $6.5 \pm 2.0$~dB for 3GPP LOS and slope-intercept prediction, respectively, where the standard deviation indicates street-by-street variation. By only using four most influential clutter features, RMSE of $5.5\pm 1.1$~dB is achieved. 
\end{abstract} 


\IEEEpeerreviewmaketitle

\section{Introduction}\label{introduction}
The 5th generation of mobile networks (5G) has adopted a much broader spectrum at higher frequency bands, such as mm-wave bands, that promises very high data rates. To unleash the full potential of mm-wave communications, highly accurate channel modeling and path loss (PL) prediction are essential in foretelling cell coverage, planning deployment of the base stations (BSs), and optimizing network performance~\cite{Phillips12}. However, high bands come with the challenge of  higher free-space, scattering, and diffraction losses from the propagation environment. For example, in a typical urban street, buildings and street clutter like scaffolding, vehicles, and tree canopies can significantly impact PL compared to lower frequency bands (wavelength of tens of cm). Although accurate PL estimation by employing fast and straightforward models is pivotal in network planning and optimization, they are yet to be fully understood in mm-wave frequencies at various propagation environments.

Numerous PL prediction models have been established in the literature, which can be classified into three major categories: statistical\cite{Erceg00}--\hspace{-0.1mm}\cite{urban1}, deterministic\cite{Sarkar03}, \cite{Montiel03}, and learning-based models\cite{Popescu06}--\hspace{-0.1mm}\cite{Ratnam21}. Statistical models such as~\cite{Erceg00} provide a computationally efficient method by fitting particular equations to measurements obtained in different propagation environments~\cite{Erceg00}--\hspace{-0.1mm}\cite{urban1}. The most widely adopted heuristic channel models, referred to as slope-intercept model hereafter, apply a linear fit to the measured PL data against the logarithm of the Euclidean distance between the transmitter and the receiver. Deterministic models such as ray-tracing, on the other hand, are based on the principles of physics to simulate wave transmission, reflection and diffraction. Its PL prediction depends not only on the environment abstraction (geometry and material properties) but also subjective parameter settings (e.g., number of rays, the maximum number of reflections). Ray tracing at the mm-wave band is especially challenging~\cite{Sarkar03, Montiel03} since the detailed characterization of foliage and street clutter and associated reflection/diffraction calculation at short wavelengths all require customized approximations (such as empirical reflection coefficients, rough surface, the inclusion of diffuse reflection and diffraction), making ``tuning'' (i.e., parameter adjustment against field measurement~\cite{Lee18, Charbonnier20, Leekim19}) an essential part of the commercial ray-tracing tools in mm-wave network planning. Therefore, machine learning (ML) based techniques have appeared as a promising alternative.

\begin{table*}
\renewcommand*{\arraystretch}{1.15}
  \begin{center}
    \caption{Comparison of learning based path loss prediction \cite{Popescu06}--\hspace{-0.1mm}\cite{Ratnam21}\vspace*{-0.1cm}} 
    \label{literature_review_table}
    {\scriptsize
    \begin{tabular}{|c|c|c|c|c|c|c|}
    \hline  
   \emph{Ref} & \emph{Environment} & \emph{Frequency} & \emph{Key Features} & \emph{ML tools} & \emph{Train/test} & \emph{Data Source} \\
   \hline
   \multirow{1}{*}{\cite{Popescu06}} & \multirow{1}{*}{Indoor} & \multirow{1}{*}{$1.89$ GHz} & Tx Position and gain, Tx height, distance  & \multirow{1}{*}{ANN} & \multirow{1}{*}{Split dataset} & \multirow{1}{*}{Measurements}\\
   \hline
  \multirow{1}{*}{ \cite{Ostlin10}} & \multirow{1}{*}{Rural} & \multirow{1}{*}{$0.881$ GHz} & Tx height, TLA, land usage& \multirow{1}{*}{ANN} & \multirow{1}{*}{Split dataset} & \multirow{1}{*}{Measurements}\\
   \hline
   \multirow{1}{*}{\cite{Monica19}} & {Desert like area} & \multirow{1}{*}{$1.8$ GHz} & \multirow{1}{*}{Terrain profile}  & {ANN, CNN} & \multirow{1}{*}{Split dataset} & \multirow{1}{*}{Measurements}\\
   \hline
   \multirow{1}{*}{\cite{Wu10}} & \multirow{1}{*}{Railway} & \multirow{1}{*}{$0.930$ GHz} & Viaducts, cuttings, plains  & \multirow{1}{*}{ANN} & \multirow{1}{*}{Split dataset} & \multirow{1}{*}{Measurements}\\
   \hline
   \multirow{1}{*}{\cite{Sotiroudis19}} & \multirow{1}{*}{Urban} & {$0.9, 1.8$ GHz} & \multirow{1}{*}{Building height, LOS path}  & {ANN, RF} & \multirow{1}{*}{Split dataset} & \multirow{1}{*}{Simulated}\\
   \hline
   \multirow{1}{*}{\cite{Ates19}} & \multirow{1}{*}{Urban} & \multirow{1}{*}{$0.9$ GHz} & \multirow{1}{*}{2D satellite images}  & {VGG-16} & \multirow{1}{*}{Split dataset} & \multirow{1}{*}{Simulated}\\
   \hline
   \multirow{1}{*}{\cite{Egi19}} & \multirow{1}{*}{Urban} & \multirow{1}{*}{$2.6$ GHz} & Tx gain, position and height, distance & \multirow{1}{*}{ANN} & \multirow{1}{*}{Split dataset} & \multirow{1}{*}{Measurements}\\
   \hline
   \multirow{1}{*}{\cite{Masood19}} & \multirow{1}{*}{Urban} & \multirow{1}{*}{$2.1$ GHz} &  \multirow{1}{*}{3D point cloud} & \multirow{1}{*}{ANN} & \multirow{1}{*}{Split dataset} & \multirow{1}{*}{Measurements}\\
   \hline
  \multirow{1}{*}{ \cite{Isabona16}} & \multirow{1}{*}{Urban} & \multirow{1}{*}{$-$} & Propagation loss, distance  & \multirow{1}{*}{ANN} & \multirow{1}{*}{Split dataset} & \multirow{1}{*}{Measurements}\\
   \hline
   \multirow{1}{*}{\cite{Park19}} & \multirow{1}{*}{Urban} & \multirow{1}{*}{$3-6$ GHz} & Distance, frequency  & \multirow{1}{*}{ANN} & \multirow{1}{*}{Split dataset} & \multirow{1}{*}{Measurements}\\
   \hline
   \multirow{1}{*}{\cite{Zhang19}} & \multirow{1}{*}{Urban} & \multirow{1}{*}{$5.8$ GHz} & Antenna separation, distance, frequency & {ANN, RF, SVR} & \multirow{1}{*}{Split dataset} & \multirow{1}{*}{Measurements}\\
   \hline
   \multirow{1}{*}{\cite{Cayamcela20}} & \multirow{1}{*}{Urban, Suburban, Rural} & \multirow{1}{*}{$2.4$ GHz} & Satellite images used for image segmentation & {AE} & \multirow{1}{*}{Split dataset} & \multirow{1}{*}{Measurements}\\
   \hline
   \multirow{1}{*}{\cite{Oroza17}} & \multirow{1}{*}{Alpine (Dense Trees)} & \multirow{1}{*}{$2.4$ GHz} & Distance, vegetation variability, terrain, canopy coverage  & {ANN, RF, kNN} & \multirow{1}{*}{Split dataset} & \multirow{1}{*}{Measurements}\\
   \hline
    \cite{Popoola18} & Urban, Rural, Suburban & $1.8$ GHz & Longitude, latitude, altitude, clutter height, elevation & ANN &  route-wise split &  Measurements\\ 
   \hline
    \multirow{2}{*}{\cite{Yuyang19}} & \multirow{2}{*}{Desert like area} & \multirow{2}{*}{$1.8$ GHz} & \multirow{2}{*}{Terrain profile}  & {RF} &  {Spatially disjoint} & \multirow{2}{*}{Measurements}\\ 
     &  &  &  & {AE} &  {split} & \\ 
   \hline
    \cite{Kuno18} &  Public square &  high frequency &  Building height, distance, LOS/NLOS & CNN & site-wise split & Simulated\\ 
   \hline
   \multirow{1}{*}{\cite{Thrane20}} & \multirow{1}{*}{Urban} & \multirow{1}{*}{$2.6$ GHz} &  \multirow{1}{*}{Satellite images}  & \multirow{1}{*}{CNN} & site-wise split & \multirow{1}{*}{Measurements}\\ 
   \hline
  \multirow{1}{*}{\cite{Gaire19}} & {Urban} &  {$5.9$ GHz} &   {2D} city map, street map,  location of cars &  {CNN} &  map-wise split    &  {Simulated}\\
	 \hline
   \multirow{1}{*}{\cite{Lee19}} & \multirow{1}{*}{Urban} & \multirow{1}{*}{$28$ GHz} &  \multirow{1}{*}{2D-collapse of 3D-buildings}  & \multirow{1}{*}{CNN} & \multirow{1}{*}{Split dataset} & \multirow{1}{*}{Simulated}\\
   \hline
   {\multirow{2}{*}{\cite{Ratnam21}}} & \multirow{2}{*}{Urban} & \multirow{2}{*}{$28$ GHz} & {Building height, terrain height,} & {CNN-based} & Split dataset \& &  \multirow{2}{*}{Simulated}\\
   & & & tree foliage height, and LOS information & U-Net & city-wise split &  \\
   \hline
    \end{tabular}}\vspace*{-0.35cm}
  \end{center}
\end{table*}

\subsection{Previous Work}
PL prediction can be considered as a regression problem in ML, where the features extracted from the propagation environment become its input and PL as a continuous variable output. We summarize some of the ML-based approaches for propagation environment modeling and PL prediction \cite{Popescu06}--\hspace{-0.1mm}\cite{Ratnam21} in Table~\ref{literature_review_table}, highlighting the propagation environment, frequency, key features, training and testing procedures, PL data source, and ML tools such as artificial neural networks (ANNs), random forest (RF), convolutional neural network (CNNs), autoencoder (AE), and support vector regression (SVR). These works showcased the capability of ML-based methods and their potential in improving PL prediction accuracy. A more comprehensive review on ML-based PL prediction can be found in~\cite{Seretis21}. 

Many of the ML-based approaches~\cite{Popescu06}--\hspace{-0.1mm}\cite{Oroza17}, \cite{Lee19} focus on prediction for nearby links (i.e., interpolation). For studies (\hspace{-0.1mm}\cite{Popoola18}--\hspace{-0.1mm}\cite{Gaire19,Ratnam21}) that do predict PL for new streets/areas (i.e., extrapolation), the influence of street clutter such as trees and street furniture on PL is either minimal or non-existent. Besides, in most previous works, complex ML models are adopted as a black-box, making it hard to interpret the connection between features and PL prediction. These complex ML models also make PL prediction vulnerable to overfitting since the training data size from measurements is usually too small compared to adjustable parameters in ML models.

Compared to studies such as \cite{Popescu06}--\hspace{-0.1mm}\cite{Gaire19} that are dedicated to the sub-$6$ GHz bands, ML-based PL prediction for mm-wave bands requires a much finer level of details in the environment description as scattering by small objects (tenths of wavelength) and material absorption loss are more significant for mm-wave signals.
For example, about $30$~dB street-by-street variation in median PL has been observed from field measurements in urban street canyons at $28$~GHz~\cite{urban1}. Recently, ML-based PL prediction at mm-wave bands in urban street canyons has been proposed using CNN-based models~\cite{Lee19,Ratnam21}. Both models are trained using ray-tracing simulated PL data and thus provide computationally efficient ways to approximate ray-tracing prediction. However, ray-tracing itself at the mm-wave band in a cluttered environment needs to be improved to better match field measurements~\cite{Lee18, Charbonnier20, Leekim19}. For example, ray tracing calculation was found to predict about 10 dB stronger received power than observed in urban street canyons at $28$~GHz~\cite{urban1}. This is thought to be due to difficulty in representing and modeling street clutter and foliage in traditional ray tracing. Therefore, accurate and practical models are still needed.

\begin{figure*}[t!]
\centering
\includegraphics[scale=0.5]{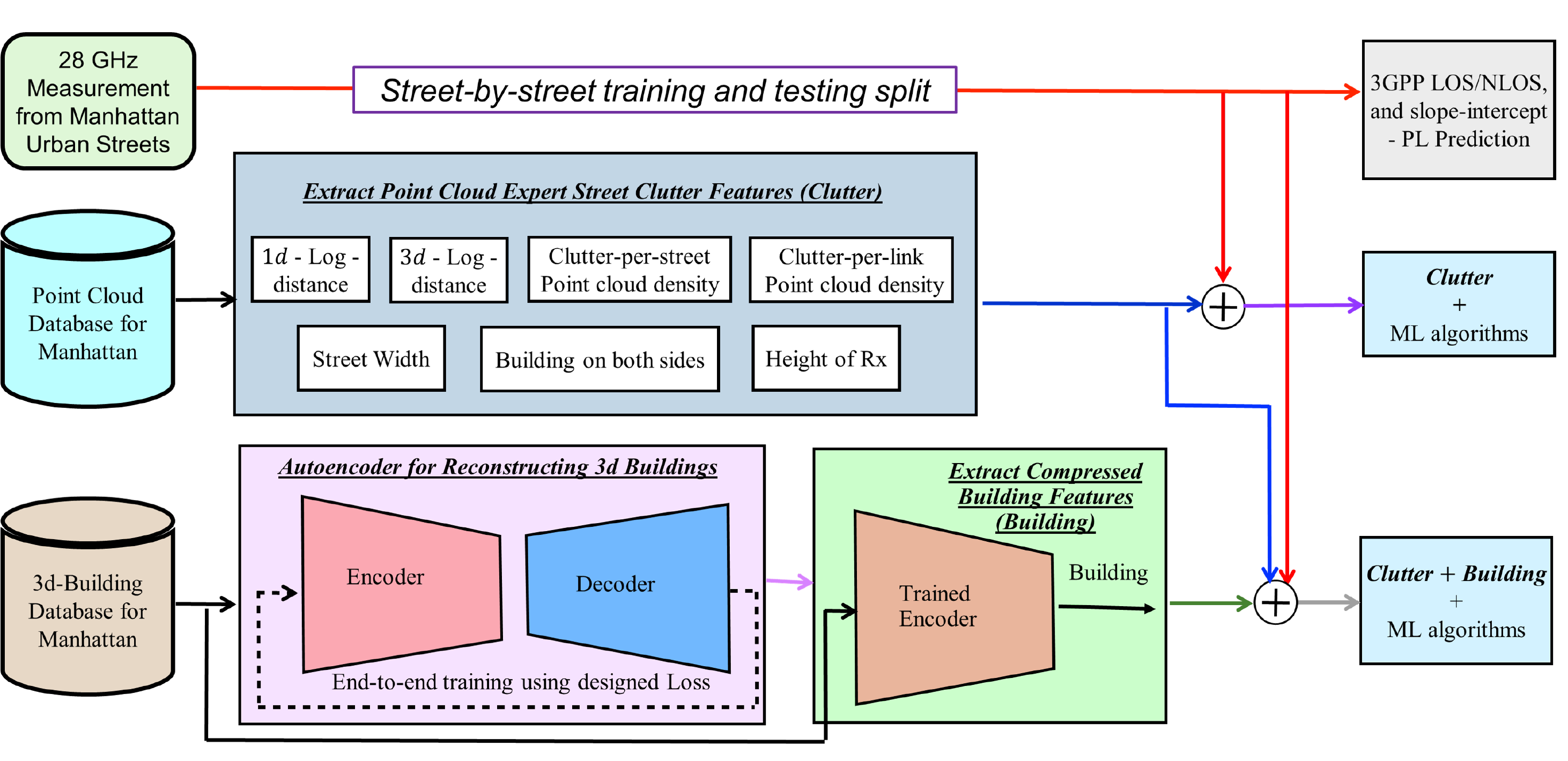}  
\caption{{Methodology adopted in this work}\vspace*{-0.25cm}.}
\label{method}
\end{figure*} 

\vspace*{-0.2cm}
\subsection{Our Contributions}\label{sec:contributions}
We address three key challenges in ML-based PL prediction for mm-wave bands: 1) \textit{reliability} due to no/insufficient measurement data; 2) \textit{generalizability} due to weak/no capability of extrapolation; 3) \textit{interpretability} due to complex ML models and high dimensional features. 

\subsubsection{Reliability} We utilize a large-scale dataset from $28$~GHz field measurements~\cite{urban1} in urban street canyons, consisting of $1028$ continuous-wave links from $13$ streets in Manhattan from multiple roof top sites co-located with commercial BS. The street clutter information such as tree canopies and lampposts   is obtained from the open-source LiDAR point cloud dataset~\cite{dataset_pc}, and the building information is obtained from the open-source 3D mesh-grid~\cite{dataset_building}, which includes building height, façade shape, separation between the buildings, roads, elevation information, etc.

\subsubsection{Generalizability} The generalizability We address it from three aspects:
\begin{itemize}
\item \emph{Street-by-street training and testing policy}: $13$ independent training-testing combinations are created by choosing one street at a time for testing and the rest $12$ streets for training. Such policy would test extrapolation of trained models to ``never seen" streets.  
\item \emph{Aggressive street clutter and building feature compression}: For each link, the high-resolution point cloud raw data is compressed to two numbers using heuristic approaches devised from expert-knowledge in wave propagation, and the 3D building information is compressed to a length-$12$ vector using CNN-based autoencoders to preserve the spatial characteristics.
\item \emph{Reducing adjustable parameters for PL prediction}: Simple ML based regression algorithms such as Lasso, Elastic-net, random forest, and SVR are adopted to mitigate over-fitting. 
\end{itemize}

\subsubsection{Interpretability} We adopt human-friendly environment features and quantify the significance of each feature in the PL prediction. We define only seven expert-knowledge driven propagation environment features, referred as  \textit{Clutter} features hereafter, where each feature has a physical meaning attached. We quantify the importance of each \textit{Clutter} feature by Lasso weight analysis and by comparing the PL prediction accuracy when only one feature is excluded. 

To the best of our knowledge, this is the first time both street clutter and building information are used collectively for mm-wave PL prediction using a large-scale real-world propagation measurement at 28 GHz in urban streets. We show that our proposed model achieves root mean square error (RMSE) of $4.8$ dB averaged over all 13 streets with $1.1$ dB  standard deviation that reflects street-by-street variation. By only using the top four most influential features, our model achieves prediction RMSE of $5.5\pm 1.1$~dB (mean$\pm$std). In contrast, the heuristic slope-intercept method and the 3GPP LOS model based prediction have RMSE of $6.5\pm2.0$ dB  and $10.6\pm4.4$ dB, respectively. For the first time, we show that the ML assisted PL predictions are more accurate than measurement based slope-intercept model with much smaller street-by-street variation. The methodology adopted in this work is summarized in Fig.~\ref{method}. 

\subsection{Paper Organization}
  The datasets of PL, point-cloud and 3D building are described in Sec.~\ref{sec:data}. Expert-knowledge based feature extraction from point cloud is presented in Sec.~\ref{sec:pointcloud} and 3D building feature compression using CNN-based AE is in Sec.~\ref{3dAE}. ML algorithms for PL prediction and the street-by-street training and testing methodology are elaborated in Sec.~\ref{PLmodels}. Performance evaluation is in Sec.~\ref{sec:results} and conclusions in Sec.~\ref{sec:conclusion}.
 
\section{PL Data Collection and Feature Sets Preprocessing}\label{sec:data}
In this section, we present an overview of the PL measurement in Manhattan and preprocessing of the point cloud dataset and the 3D building mesh grid dataset.

\begin{figure}[t!]
\centering
\begin{subfigure}[t]{0.48\textwidth}
\centering\hspace*{-0.35cm}
\includegraphics[scale=0.41]{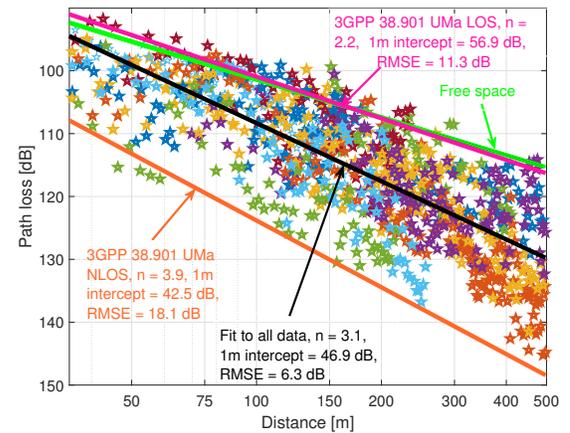} 
\caption{ {Measured path loss of $1028$ links from $13$ streets}\vspace*{-0cm}}
\label{intro_1}
\end{subfigure}
\begin{subfigure}[t]{0.48\textwidth}
\centering\hspace*{-0.35cm}
\includegraphics[scale=0.41]{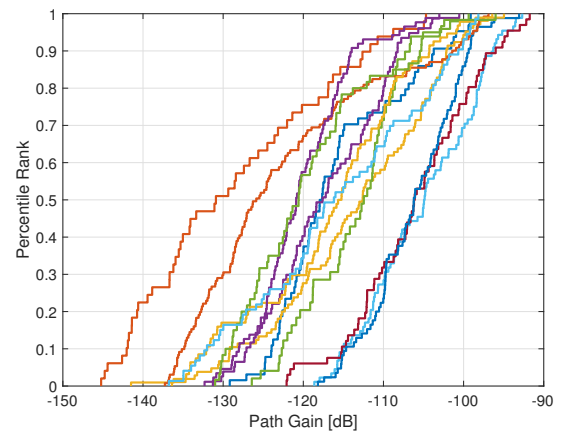}
\caption{{Large street-by-street variation}\vspace*{-0.05cm}}
\label{intro_2}
\end{subfigure}
\caption{\label{intro}Manhattan measurement data and street-by-street variation (different color per street).\vspace*{-0.5cm}
}
\end{figure}

\subsection{28 GHz PL Measurement in Manhattan}\label{sec:measured}
The measurement campaign~\cite{urban1} is designed to characterize coverage in an urban street canyon, which can be coarsely defined as a straight road in an urban area that has buildings on both sides, with street clutter such as trees, vehicles, and lampposts placed along the road. Measurement was done from roof edge mounted BS (i.e., urban macro) to UEs ($1.5$ m high) in the center of a sidewalk along the street, with no attempt to incorporate or eliminate blockage due to street clutter. The purpose is to resemble coverage of the street in the presence of such obstructions. 

\begin{table*}
\renewcommand*{\arraystretch}{1.15}
  \begin{center}
    \caption{Diverse characteristics of 13 measured streets and their Point Cloud-based {Clutter} features.}
    \label{expert_feat_table}
    {\footnotesize
    \begin{tabular}{|c||c|c|c|c|c|c|c|c|c|}
    \hline  
   \emph{Street} & {\emph{Building height}}  & {\emph{Terrain}} & {$\#$\emph{measured}} & {\emph{{1D}- distance}} &{Clutter-per} & {Clutter-per} & \emph{Street} & \emph{Building on} & \emph{Height}  \\ 
   \emph{Number} & {\emph{range}} & {\emph{variation}} & \emph{links} & {\emph{range}} & {-street} & {-link range} &  \emph{width} & \emph{both sides} & \emph{Rx}  \\
   \hline \hline
   $1$ & $23-69$ m & $0-3$ m & $88$  & $32-165$ m & $0.51$ & $0-7$  & $25$ m & \Checkmark & $15$ m \\
   \hline 
   $2$ & $11-43$ m  & $<1$ m & $77$  & $69-338$ m & $2.71$ & $0 - 59$  & $23$ m & \Checkmark & $35$ m  \\
   \hline 
   $3$ & $14-72$ m & $0-24$ m & $94$  & $36-492$ m & $1.86$ & $0-14$ & $28$ m & \Checkmark & $15$ m \\
   \hline 
   $4$ & $12-61$ m & $<1$ m & $105$  &  $35-453$ m & $2.74$ & $0-26$  & $36$ m & \Checkmark & $25$ m \\
   \hline 
   $5$ & $16-67$ m & $0-3$ m & $85$  & $32-163$ m & $0.51$ & $0-8$  & $25$ m & \Checkmark & $15$ m  \\
   \hline 
   $6$ & $14-72$ m & $0-23$ m & $131$  & $34-495$ m & $1.86$ & $0-17$  & $28$ m & \Checkmark & $15$ m  \\
   \hline 
   $7$ & $12-30$ m & $<1$ m & $87$  & $153-497$ m & $1.70$ & $0-19$  & $15$ m & \Checkmark & $20$ m \\
   \hline 
   $8$ & $14-93$ m & $<1$ m & $64$  & $92-490$ m & $0.32$ & $0-1$ & $30$ m & \Checkmark & $22$ m \\
   \hline 
   $9$ &$16-67$ m  & $<1$ m & $66$ & $36-235$ m & $1.51$ & $0-20$  & $27$ m & \Checkmark & $15$ m \\
   \hline 
  $10$ & $22-66$ m & $<1$ m & $73$  & $37-217$ m & $3.42$ & $0-13$  & $27$ m & \xmark & $40$ m  \\
   \hline 
  $11$ & $15-74$ m  & $<1$ m & $49$  & $151-411$ m & $0.84$ & $0-20$ & $38$ m & \Checkmark & {$54$ m} \\
   \hline 
  $12$ & $15-52$ m & $<1$ m & $49$  & $182-497$ m & $5.04$ & $0-84$  & $18$ m & \Checkmark & $22$ m\\
   \hline 
  $13$ & $30-70$ m & $<1$ m & $63$  & $36-255$ m & $4.18$  &  $0-16$ & $27$ m & \xmark & $40$ m  \\
   \hline\hline
	{ Overall} & $11-93$ m & $0-24$ m & $49-131$  & $32-497$ m & $0.51{-}5.04$  & $0-84$  & $15{-}38$ m & 11 Yes, 2 No & $15{-}54$ m  \\
	\hline
   \end{tabular}}
  \end{center}\vspace{-0.25cm}
\end{table*}

\begin{figure*}[t!]
\centering\hspace*{-0.5cm}
\begin{subfigure}[t]{0.25\linewidth}
\centering
\includegraphics[scale=0.6]{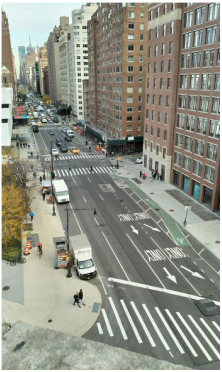}
\caption{$7^\mathrm{th}$ Avenue, Manhattan}
\label{Aerial_1}
\end{subfigure}
\begin{subfigure}[t]{0.25\linewidth}
\centering
\includegraphics[scale=0.6]{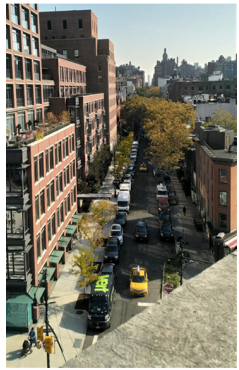}
\caption{W $11^\mathrm{th}$ Street, Manhattan}
\label{Aerial_2}
\end{subfigure}
\begin{subfigure}[t]{0.45\linewidth}
\centering
\includegraphics[scale=0.37]{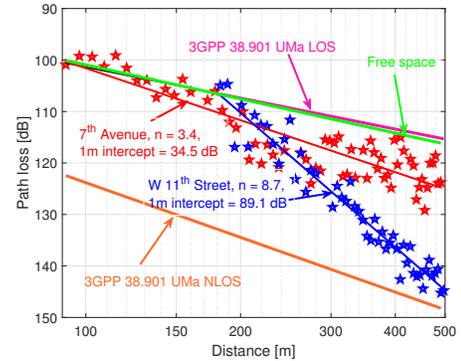}\vspace{-0.35cm}
\caption{{Comparison of measured PL}\vspace{-0.75cm}}
\label{pc_0}
\end{subfigure}\\\hspace*{-0.5cm}
\begin{subfigure}[t]{0.45\linewidth}
\centering\hspace*{-0.5cm}
\includegraphics[scale=0.18]{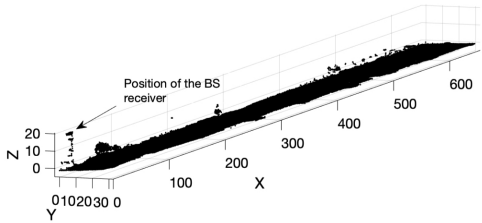}
\caption{Point cloud for $7^\mathrm{th}$ Avenue.}
\label{pc_1}
\end{subfigure}
\begin{subfigure}[t]{0.45\linewidth}
\centering
\includegraphics[scale=0.18]{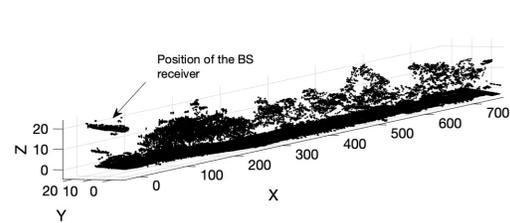}
\caption{Point cloud  for W $11^\mathrm{th}$ Street.}
\label{pc_2}
\end{subfigure}
\caption{Two streets covered from the same roof top with diverse street clutter density.\vspace{-0.25cm}}
\end{figure*}

Measurements were performed from $7$ building rooftops covering $13$ streets from multiple areas of Manhattan of different street width and widely varying amount of foliage, from the \emph{Pike Street} in the Lower East Side to \emph{W 126th Street} in West Harlem. The propagation environment  differs significantly among the $13$ streets. For example, street width ranges from 15 to 38 m whereas building height varies from 11 to 93 m. Tree distribution along streets ranges from nothing to sparse, to very dense, and some streets even have road dividers with trees/bushes separating driveways. We summarize the diverse characteristics of the 13 streets in Table~\ref{expert_feat_table}\footnote{Details of Clutter-per-street and Clutter-per-link are deferred to Sec.~\ref{features_compression}.}.

 In total, $1028$ links were measured  from over $3,800$ m of street side walks with over $10$ million individual power measurements, which were locally averaged per link to eliminate small scale fading.
 In Fig. \ref{intro_1} we show the measured links and the slope-intercept fit to PL versus logarithmic Euclidean distance $(d)$, given by
\begin{align}
P  = A + 10n\log_{10}(d) + \mathcal{N}(0, \sigma^2), \label{conv_SI_model}
\end{align}
where $A {=} 46.9$ dB is the $1$ m-intercept, {$n {=} 3.1$} represents the slope, $\sigma {=} 6.3$ dB is the RMSE between fitted and actual values and $\mathcal{N}(\cdot)$ is the normal distribution representing shadow fading. Comparing our data against standard PL models such as 3GPP UMa LOS and NLOS~\cite{38.901} leads to RMSE of $11.3$ dB and $18.1$ dB, respectively. Fig. \ref{intro_2} depicts the distributions of path gain\footnote{Path gain, instead of path loss, is displayed in Fig.~\ref{intro_2} to emphasize the deteriorating link quality at lower percentile rank.} for individual streets, with median spanning over a range about $30$ dB. 

The presence of street clutter may be the cause for about $10$~dB excess loss compared to the UMa LOS model in Fig. \ref{intro_1} and the large street-by-street variations in Fig.~\ref{intro_2}. This is due to the short wavelengths of mm-wave signals, approximately $1$~cm at $28$~GHz, making them more susceptible to intense diffused scattering and poorer rough surface reflection~\cite{review_ref2}. Further, mm-wave has a tighter first Fresnel zone directly proportional to its wavelength, causing objects as small as tens of centimeters to appear to be substantial in impairing link quality~\cite{review_ref4}. This motivates us to capture detailed environment features such as street clutter and 3D-building. 

\subsection{Street Clutter Modeling using LiDAR Point Cloud Dataset}
Let us consider the measurements done from the same rooftop for two Manhattan streets, the $7^\mathrm{th}$ Avenue with a handful of young trees and the W $11^\mathrm{th}$ Street with many tall tree canopies,  as shown in Fig. \ref{Aerial_1}, and Fig. \ref{Aerial_2}, respectively. Measured PL and their slope-intercept fits are shown in Fig. \ref{pc_0}. The distance exponent of the W $11^\mathrm{th}$ Street is significantly higher, $8.7$, compared to $3.4$ on the $7^\mathrm{th}$ Avenue, with a $23$~dB gap in average PL at $500$ m. Therefore, street clutter information, which includes tree canopies, cars, lampposts, etc., plays a crucial role in PL prediction.

To capture the street clutter, we use the \emph{USGS CMGP LiDAR} point cloud repository~\cite{dataset_pc}, where each object is described by a set of points on its external surfaces acquired at 1 cm resolution.
For each street, we change the origin to the ground location of the Rx position and align $X$-axis with the street along which the Tx is moving, $Y$-axis along the width of the street, and $Z$-axis pointing to the Rx placed at the top of the building. We also utilize a \textit{k-nearest neighbor} based point cloud denoising~\cite{Rusu08}. The processed point cloud representing the street clutter for the $7^\mathrm{th}$ Avenue and the W $11^\mathrm{th}$ Street are shown in Fig.~\ref{pc_1}, and Fig.~\ref{pc_2}, respectively.  
 
\subsection{3D-Building Mesh Grid}\label{2dccollapsesection}
\begin{figure}[t!]
\centering
\includegraphics[scale=0.537]{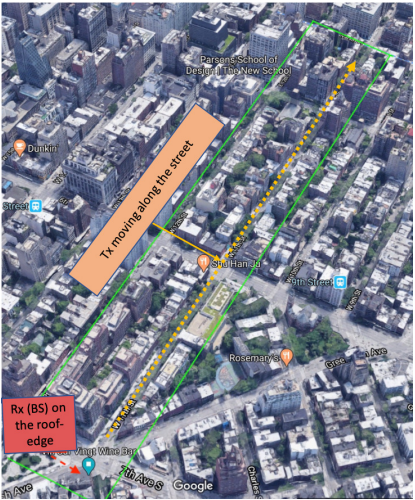}
\caption{{Aerial-view of W $11^\mathrm{th}$ Street}.\vspace{-0.25cm}}
\label{build_1}
\end{figure}

\begin{figure}[t!]
\centering\hspace*{-0.26cm}
\begin{subfigure}[t]{\linewidth}
\centering
\includegraphics[scale=0.53]{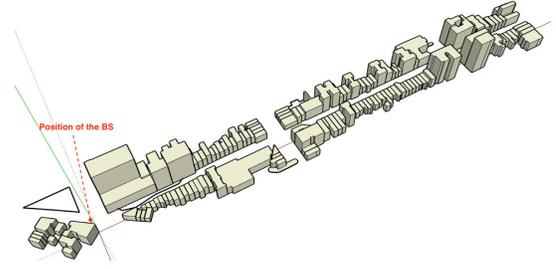}
\caption{{3D building (mesh grid) acquired for the W $11^\mathrm{th}$ Street.}}
\label{build_2}
\end{subfigure}\\
\begin{subfigure}[t]{\linewidth}
\centering\hspace*{-0.25cm}
\includegraphics[scale=0.225]{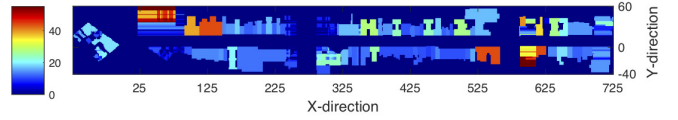}
\caption{{{2D} collapse for the W $11^\mathrm{th}$ Street mesh grid buildings.}}
\label{build_3}
\end{subfigure}
\caption{{Example of 3D building mesh grid data.}\vspace{-0.25cm}}
\end{figure} 
 
Reflection and scattering from urban buildings can be significant and impose wave-guiding effect on the signals. We extract the 3D building mesh grid from the Cadmapper~\cite{dataset_building}. The aerial-view of the W $11^\mathrm{th}$ Street is shown in Fig.~\ref{build_1} and the extracted 3D buildings are in Fig.~\ref{build_2}. We first convert the high dimensional mesh grid into a Euclidean space by assigning  each $1\times1\times1$~m cube a value 1/0 indicating the presence/absence of a mesh grid. We then reposition the origin such that Rx is at $(0, 0, \text{Rx height})$ and align the $X$-axis along the street and $Y$-axis along street width (i.e., align the coordinates with those used for point cloud). To reduce the dimensionality of the dataset while preserving height information, we collapse it along the $Z$-axis into $2$D grids (of $1 \times 1$ m) and assign the entry of each grid the height of building at that location ($0$ if there is no building), as shown in Fig.~\ref{build_3} for the W $11^\mathrm{th}$  Street where the color bar indicates height. 

\begin{remark}
Please note that accurate Point-Cloud and 3D-meshgrid datasets collected along with the PL measurements are desirable but that would significantly increase the cost and overhead. Since we are interested in capturing the general characteristics of the street clutter and 3D buildings rather than the fine details, we find it is sufficient to use publicly available Point-Cloud and 3D-meshgrid datasets from \cite{dataset_pc} and \cite{dataset_building}, respectively, independent from the PL measurements~\cite{urban1}.  
\end{remark}

\section{{Street Clutter Feature Extraction} from Point Cloud}\label{sec:pointcloud} 
The LiDAR point cloud dataset~\cite{dataset_pc} contains massive amount of data that can't be directly used for training or interpretation. Thus, we focus on expert-knowledge driven feature extraction from the point cloud dataset for modeling the street clutter.
	
\subsection{Street Clutter Feature Compression}\label{features_compression}
We compress all the LiDAR point cloud information for each link into two numbers using heuristic approaches devised from expert-knowledge in wave propagation, where each number is proportionally to the count of points in a 3D volume:

\begin{itemize}
\item \emph{Street-specific clutter density (Clutter-per-street)} --  averaged point cloud density in a 3D volume defined by (along-street distance between Rx and \textbf{furthest} Tx) $\times$ (street width) $\times$ (Rx height above ground). 
\item \emph{Link-specific clutter density (Clutter-per-link)} -- total number of points contained in a string of contiguous $1\times 1\times 1$~m cubes traversed by the straight line connecting Tx and Rx.
\end{itemize}

\begin{figure}
\centering\hspace*{0.2cm}
\includegraphics[scale=0.25]{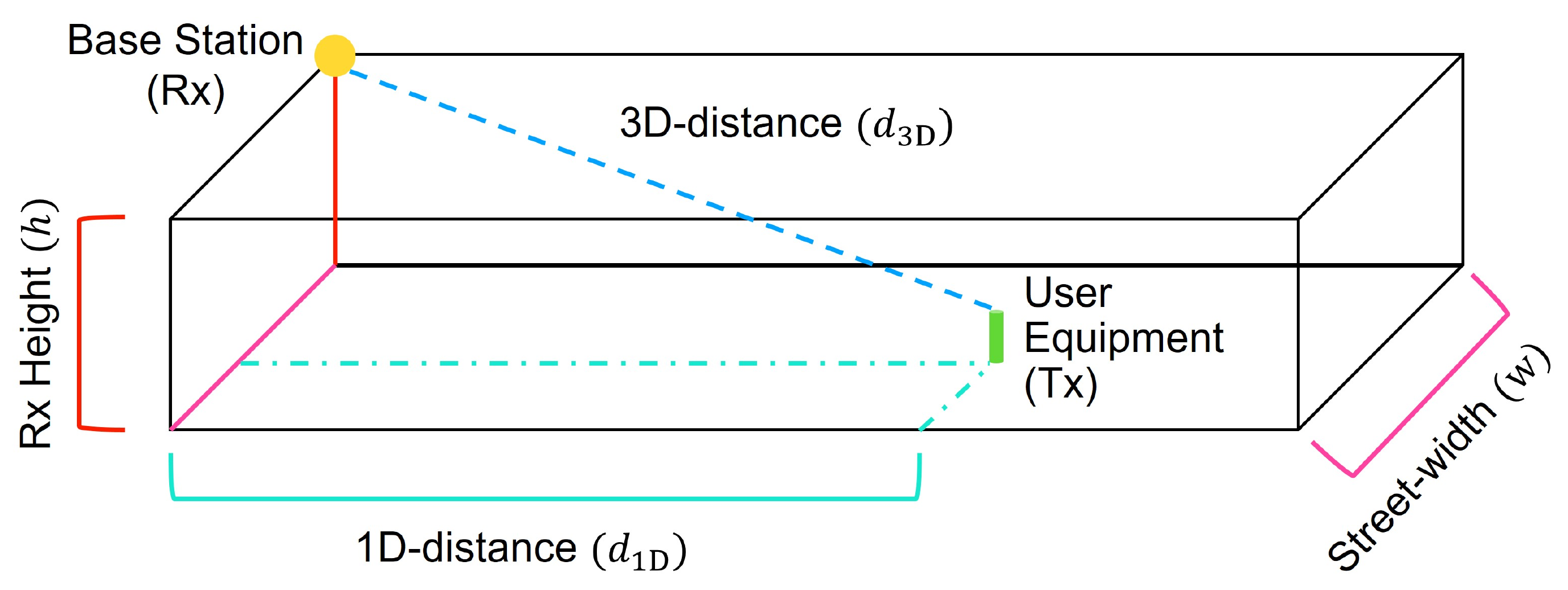}
\caption{Illustration of Point Cloud based Clutter features.\vspace{-0.35cm}} \label{Clutter_explain_diag_label}
\end{figure}
\begin{figure*}[t!]
\centering
\includegraphics[scale=0.35]{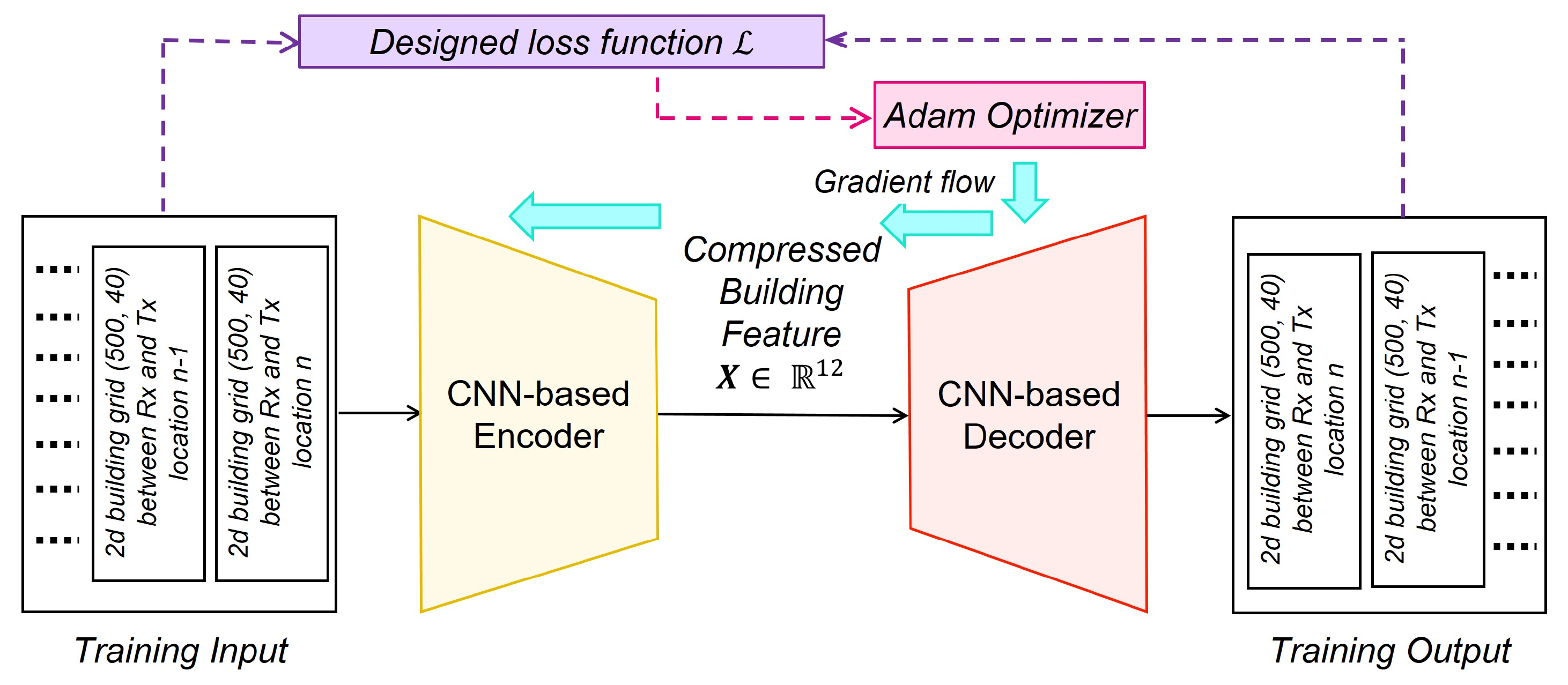}\vspace*{-0.15cm}
\caption{{Proposed CNN-based autoencoder for building feature compression.\vspace*{-0.15cm}}}
\label{dl_arch}
\end{figure*}

The \emph{{Clutter-per-link}} represents the {accumulated} clutter density along the direct path within the first Fresnel zone (about or smaller than the $1$ m $\times 1$ m cross-section) and thus may be interpreted as a blockage indicator of the direct path for each link. The \emph{Clutter-per-street} represents the overall clutter density of the entire street and remains the same for all links from the same street.  

\subsection{Point Cloud based Expert {Street Clutter Features (Clutter)}}\label{expert_features_section} 
The following seven expert {street clutter} features are {defined} for PL prediction:
\begin{itemize}
\item \emph{Clutter 1: log-3D distance} - Euclidean distance (log-scale) between the Tx and Rx.
\item \emph{Clutter 2: log-1D distance} - Along-the-street distance (log-scale) between the Tx and Rx.
\item \emph{Clutter 3, 4: {Clutter-per-street} and {Clutter-per-link}} - Clutter density information as defined in Sec.~\ref{features_compression}.
\item \emph{Clutter 5: Street width} - Spans from $15$ to $38$ m.
\item \emph{Clutter 6: Buildings on both sides} - Indication of guiding effect from street canyon.
\item \emph{Clutter 7: Rx height} -  Spans from $15$ to $54$ m. 
\end{itemize}

Among the defined seven Clutter features, four of them have been used in different 3GPP models: 1D distance $(d_{1\text{D}})$, 3D distance $(d_{3\text{D}})$, street width, and the base station or Rx height $(h)$, as shown in Fig.~\ref{Clutter_explain_diag_label}. Such features  provide us a way to interpret the trained models and compare them against the 3GPP and slope-intercept models. The other three features capture street-specific (\emph{Clutter-per-street}) and link-specific (\emph{Clutter-per-link}) clutter information as well as the street canyon information (building on both sides of the street).

We summarize the defined expert features in Table~\ref{expert_feat_table}, wherein we omit the details of \emph{Log-3D distance}, \emph{Log-1D distance} and \emph{Clutter-per-link} features because they have a separate entry for each link. We also report the range of building height, terrain variation, and the number of measurement links on each street. Please note that the 10 times variation in the \emph{Clutter-per-street} feature ($0.51$ to $5.04$) indicates the diverse nature of street clutter in the measurement streets, consistent with the $30$~dB street-by-street variation seen in Fig.~\ref{intro_2}. 

We standardize the defined Clutter features before training and testing: for feature $f$ we compute its mean $\mu$ and variance $\sigma^2$ on the training dataset; then, we rescale the feature in both the training and testing dataset as $\hat{f} = (f-\mu)/\sigma$.

\begin{figure*}[t!]
\centering
\begin{subfigure}[t]{\textwidth}
\centering
\includegraphics[scale=0.4]{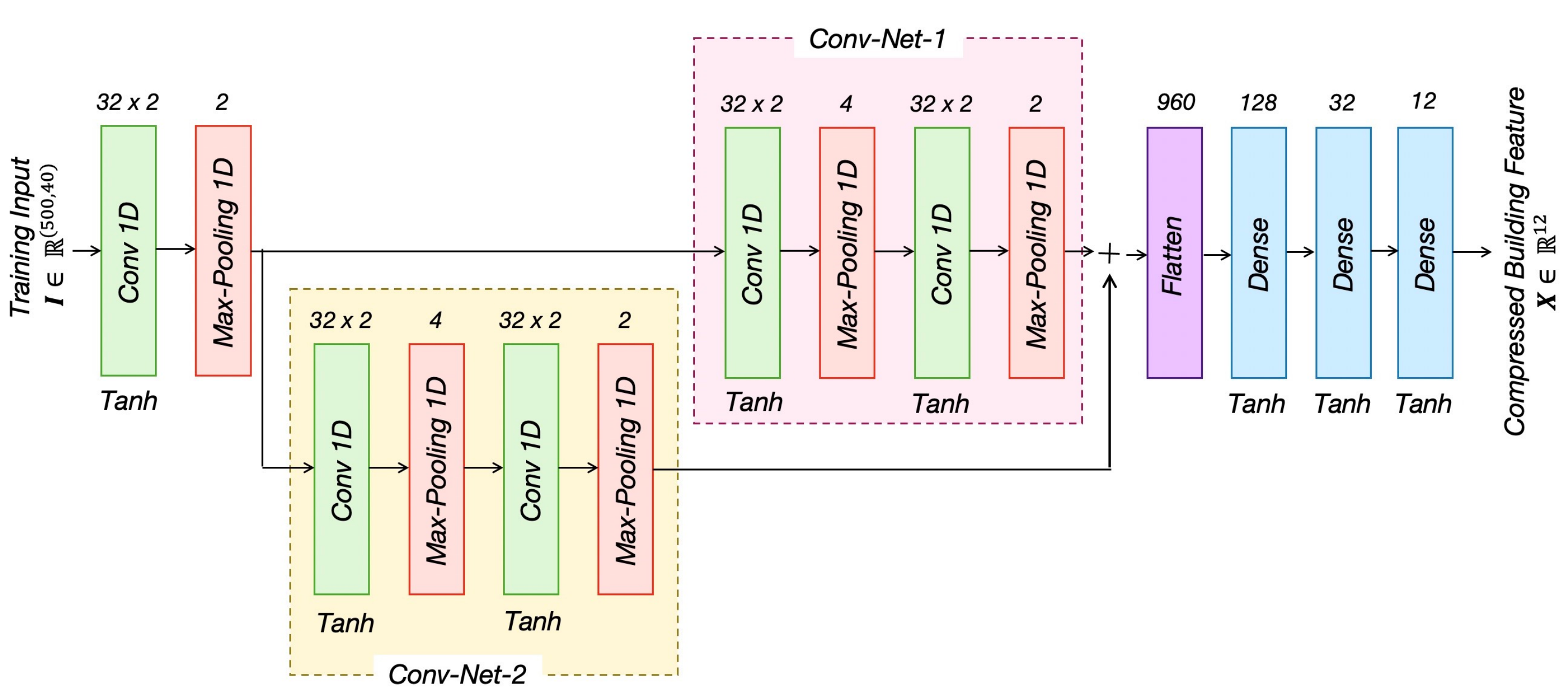}\vspace*{-0.05cm}
\caption{{Architecture of CNN-based encoder of AE.}}\label{dl_arch_1}
\end{subfigure}
\begin{subfigure}[t]{\textwidth}
\centering
\includegraphics[scale=0.4]{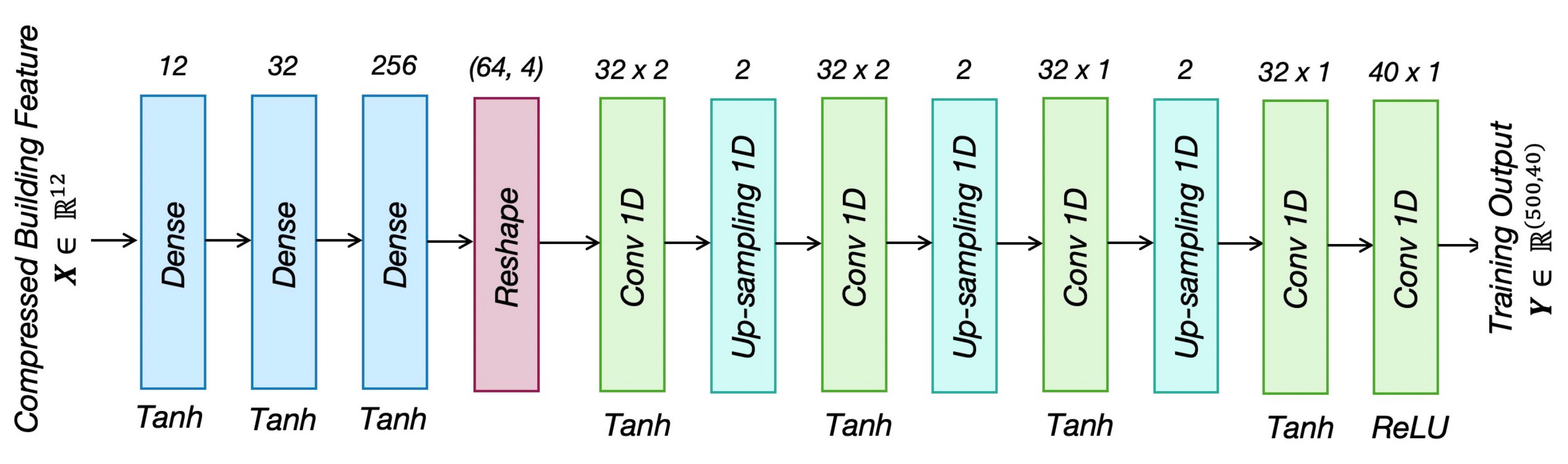}\vspace*{-0.05cm}
\caption{Architecture of CNN-based decoder of AE.\vspace*{-0.15cm}}\label{dl_arch_2}
\end{subfigure}
\caption{\label{dl_arch_0}Neural network architectures for the proposed CNN-based autoencoder. \vspace{-0.35cm}}
\end{figure*}

\section{Autoencoder based Feature Extraction from Building Dataset}\label{3dAE}
Although we have converted the $3$D building mesh grid to a 2D matrix representation, as explained in Sec.~\ref{2dccollapsesection}, the building features are still much richer than the PL data. We further reduce the 2D collapse of buildings to size of $(500, 40)$ by removing buildings beyond the maximum measurement distance $500$ m and by including the building façade only (taking grids $20$ m from the center of the street on each side). As the Tx-Rx distance increases, the number of 2D grids in between the two also increases. We append zero to the 2D matrix representation at shorter distances to preserve the distance dependency and avoid information loss by downsampling. 

We then use CNN based AE to compress the extensive building façade data to a few most relevant features for PL prediction. CNN captures the spatial dependencies with the {help} of kernels and filters, and AE learns an efficient encoder (for feature compression) and a matching decoder (to reproduce the original input signal) in an unsupervised manner. Therefore, a combination of CNN and AE, as shown in Fig.~\ref{dl_arch}, can help us to reduce the massive feature dimension of the building data to $12$ features while preserving the spatial characteristics.

The encoder in the CNN-based AE, as shown in Fig.~\ref{dl_arch_1}, takes 2D-buildings $\mathbf{I} \in \mathbb{R}^{(500, 40)}$ as input and reduces it to compressed features $\mathbf{X} \in \mathbb{R}^{(12)}$. This is achieved by the convolutional and max-pooling layers that help to reduce the dimension while preserving the spatial characteristics, whereas the dense layers extract compressed non-linear features. After the first max-pooling layer we perform grouped convolutions -- convolutions in parallel, wherein two identical CNNs (Conv-Net-1 and Conv-Net-2) are processed in parallel before their addition.

Then the decoder in the CNN-based AE, {as shown in Fig.~\ref{dl_arch_2}}, takes $\mathbf{X}$ as its input to reconstruct the 2D buildings {$\textbf{Y}\in\mathbb{R}^{(500,40)}$. This is achieved by} the dense layers that decode information from the compressed feature representation, and the convolutional and upsampling layers that revert the building information to its original form.

We design a loss function $\mathcal{L}(\cdot)$ based on the Log-cosh loss instead of MSE to increase its robustness against outliers in the building data and also reduces the impact of appended zeros in the input $\mathbf{I}$. Note that the values in $2$D matrix representation is standardized between $0$ and $1$ before feeding to the AE. For a grid $(a,b)$, we find the maximum and minimum values for that grid in all the training data and denote it by $\max(a,b)$ and $\min(a,b)$. Then we rescale all the values in the training and testing data for the $(a,b)^\mathrm{th}$ grid as $(\text{value in (a,b)} - \min(a,b))/(\max(a,b) - \min(a,b))$. The details of the designed loss function $\mathcal{L}(\cdot)$ and the CNN-based AE architecture can be found in Appendix~\ref{appenda}. 

We implement the AE in Keras~\cite{keras} with TensorFlow as a backend. We keep the learning rate of $0.0012$, batch size $16$, and the total number of epochs  $100$. We train the network end-to-end  using the Adam optimizer \cite{adam} over the time, to reconstruct the input 2D collapse of the building at the output of the decoder. Once the AE is converged\footnote{The designed AE with parallel Conv-nets (CNNs) converges after $50$ epochs and has better reproducibility than using a single or two serially concatenated Conv-nets, shown in Appendix~\ref{AE_convergence_appendix}.}, we utilize the encoder to design compressed \textit{Building} features $\mathbf{X} \in \mathbb{R}^{(12)}$ and then feed them to PL prediction.

\section{ML-based Models and Training-Testing Methodology} \label{PLmodels}
We utilize the extracted \textit{Clutter} and compressed \textit{Building} features and compare the following regularized-linear and non-linear ML algorithms~\cite{DLbook} for PL prediction using a street-by-street training and testing methodology to emphasize generalizability.

\begin{algorithm*}[t!]
\caption{Links-shuffle-split training and testing procedure}\label{alg:conv}
{\small
{
\begin{algorithmic}[1]
\Require $\boldsymbol{\mathcal{F}_{(\cdot)}}$ and $\mathbf{P}$ for $1028$ PL measurements. \Comment{Collection of all the measurements from 13 streets}.
\Ensure $\boldsymbol{\mathcal{D}} = [\mathbf{x}:=\boldsymbol{\mathcal{F}_{(\cdot)}}, \mathbf{y}:=\mathbf{P}]$. \Comment{This symbolises our total dataset}.
\For{$i=1$ \textbf{to} $iterations$} \Comment{Loop for varying shuffle and split of dataset.}
   \State Randomly shuffle and split $\boldsymbol{\mathcal{D}}$ in $4:1$ ratio to form $[\mathbf{x}_{\text{train}}^i,\mathbf{y}_{\text{train}}^i]$ and $[\mathbf{x}_{\text{test}}^i,\mathbf{y}_{\text{test}}^i]$.
   \State Train ML-based PL models (detailed in Section~\ref{ML_model}), using $[\mathbf{x}_{\text{train}}^i,\mathbf{y}_{\text{train}}^i]$ to obtain trained model $\mathcal{M}^i$.
   \State Test model $\mathcal{M}^i$ using $\mathbf{x}_{\text{test}}^i$ to predict PL $\mathbf{y}_{\text{pred}}^i$.
   \State Calculate $RMSE^i$ in PL between ground-truth PL $\mathbf{y}_{\text{test}}^i$ and predicted PL $\mathbf{y}_{\text{pred}}^i$.
\EndFor
\State {Mean RMSE $:=\mu(RMSE^{\{1,2,...,iterations\}})$ and std. deviation $:=\sigma(RMSE^{\{1,2,...,iterations\}})$.}
\end{algorithmic}}}
\end{algorithm*}

\begin{algorithm*}[t!]
\caption{Street-by-street training and testing procedure}\label{alg:prop}
{\small
\begin{algorithmic}[1]
\Require $\boldsymbol{\mathcal{F}_{(\cdot)}}^{\{1, ..., 13\}}$ and $\mathbf{P}^{\{1, ..., 13\}}$ for each of the $13$ streets.  
\For{$k=1$ \textbf{to} $13$} \Comment{Loop for each of the 13 streets as testing.}
   \State Testing Street $:= k^\mathrm{th}$ street, Training streets $:= \text{All the 13 streets \textbf{except}} \;k^\mathrm{th}$ street $:=\{1, ..., 13\}-\{k\}$.
   \State $\mathbf{x}_{\text{test}} := \boldsymbol{\mathcal{F}_{(\cdot)}}^{k}$, $\mathbf{x}_{\text{train}} := \boldsymbol{\mathcal{F}_{(\cdot)}}^{\{1, ..., 13\}-\{k\}}$, $\mathbf{y}_{\text{test}} := \mathbf{P}^{k}$ and $\mathbf{y}_{\text{train}} := \mathbf{P}^{\{1, ..., 13\}-\{k\}}$ \Comment{Measurements of $k^\mathrm{th}$ street becomes testing data, while all the other measurements form training data.}
   \State Train ML-based PL models (detailed in Section~\ref{ML_model}), using $[\mathbf{x}_{\text{train}},\mathbf{y}_{\text{train}}]$ to obtain trained model $\mathcal{M}^k$.
   \State Test model $\mathcal{M}^k$ using $\mathbf{x}_{\text{test}}$ to predict PL $\mathbf{y}_{\text{pred}}$.
   \State Calculate $RMSE^k$ in PL between ground-truth PL $\mathbf{y}_{\text{test}}$ and predicted PL $\mathbf{y}_{\text{pred}}$ for $k^\mathrm{th}$ testing street.
\EndFor
\State Mean RMSE $:=\mu(RMSE^{\{1,2,...,13\}})$ and  std. deviation $:=\sigma(RMSE^{\{1,2,...,13\}})$.
\end{algorithmic}}
\end{algorithm*}

\subsection{ML-based models for PL prediction}\label{ML_model}
Let $P$ denotes the true PL value, $\mathcal{F}_{(\cdot)}$ represents the input feature vector, and $w$ indicates the designed weight matrix. The following ML algorithms are used for PL prediction.
\begin{enumerate}
\item \textit{Lasso regression (Lasso)} optimizes the regression weights by minimizing the least-square error including a supplementary $l_1$-norm penalty on the regression coefficients (weights),
\begin{equation}
\min_w\qquad({1}/{2n_{\text{samples}}})\times\vert\vert \mathcal{F}_{(\cdot)}  w - P \vert\vert_2^2 + \alpha\vert\vert w\vert\vert_1
\end{equation} 
where $n_{\text{samples}}$ is the sample size and $\alpha>0$ imposes the $l_1$ penalty on the weights.  
\item \textit{Elastic-net regression (Elastic-net)} imposes both $l_1$- and $l_2$-norm penalties on the weights, where convex combination of $l_1,l_2$ penalties is controlled by parameter $\delta$,  
\begin{align}
\min_w\qquad &({1}/{2n_{\text{samples}}})\times\vert\vert \mathcal{F}_{(\cdot)}w - P \vert\vert_2^2 + \alpha\delta\vert\vert w\vert\vert_1 \nonumber\\
&\qquad + ({\alpha(1-\delta)}/{2})\times\vert\vert w\vert\vert_2^2
\end{align}
\item \textit{Random forest (RF)} is an ensemble learning method, where multiple decision trees' average is utilized to predict the PL. We consider $20$ estimators with maximum tree depth of $25$.
\item \textit{Support vector regression (SVR)} solves the following primal problem
\begin{align}
\min_{w, b, \zeta}\qquad &(w^Tw)/{2}  + C\sum\nolimits_{n=1}^T\zeta_n\\
\text{s.t.}\qquad &P_n(w^T\phi(\mathcal{F}_{(\cdot)_n})\!+\!b) \geq 1\!-\!\zeta_n,\nonumber\\
&\zeta_n \geq 0, \qquad \forall\;n\in[1,T]\nonumber
\end{align}
where $C$ denotes the penalty term, $\zeta_n$ indicates the distance of $n^\mathrm{th}$ sample from the decision boundary, $b$ represents the bias term and $\phi(\mathcal{F}_{(\cdot)_n})$ maps $\mathcal{F}_{(\cdot)_n}$ to a higher dimensional space. We train the SVR with an radial bias function (RBF) kernel, given by $\exp(-\gamma||\mathcal{F}_{(\cdot)_i}-\mathcal{F}_{(\cdot)_j}||^2)$ for any two samples $i$ and $j$, and $\gamma>0$.
\end{enumerate}
All the PL prediction methods are implemented using scikit-learn~\cite{scikit}. We use grid search with 5-fold cross validation~\cite{DLbook} over the training set to obtain the best parameters. In particular, the hyper-parameter $\alpha$ in Lasso and Elastic-net and $C, \gamma$ in SVR is grid-searched from $\{10^{-4}, 10^{-3}, ..., 10^{3}, 10^{4}\}$ during the training and best fitted-parameter is used for testing.

\subsection{Feature vectors}\label{sec:features}
The proposed ML-based PL prediction models can be implemented using either of the following descriptive features:

\subsubsection{{Point cloud-based expert street clutter features (Clutter) only}}\label{fepolicy1}
It consists of the {seven expert} features extracted from the street clutter information (in Sec.~\ref{expert_features_section}), represented as
\begin{align}
\mathcal{F}_{\text{Clutter}}  = &  \left\lbrace \text{\emph{log-3D distance, log-1D distance, street width,}}\right. \nonumber\\
&\quad \left. \text{\emph{{Clutter-per-link},  {Clutter-per-street}, Rx height, }}\right. \nonumber\\
&\qquad\left. \text{ \emph{buildings on both sides}} \right\rbrace \label{LRFSeq}
\end{align}

\subsubsection{Combination of {{Clutter} and Building Features (Clutter {+} Building)}} \label{fepolicy3}
Herein we first use the encoder of the trained AE (proposed in Sec.~\ref{3dAE}) to obtain compressed features $\mathbf{X} \in \mathbb{R}^{(12)}$, and concatenate with the {Clutter} in \eqref{LRFSeq}, denoted by $\mathcal{F}_{\text{Clutter\_Building}} = \left[\mathcal{F}_{\text{Clutter}}, \mathbf{X}\right]$. 

\subsection{Training and Testing Methodology}\label{testproc} 
In conventional ML-based training-testing, the $1028$ PL measurements (collection of all the measurements from $13$ streets) dataset is randomly shuffled and divided into $4:1$ ratio for training and testing sets. We refer to it as \emph{links-shuffle-split training and testing}, briefly described in Algorithm \ref{alg:conv}. To capture the impact of random shuffling and splitting, we perform the process multiple times and obtain mean RMSE and standard deviation in RMSE due to random  shuffling and splitting of the dataset. Data in the testing set is statistically similar to those in the training set, and the focus of trained models is on interpolation. Since links that are close to each other  have similar PL values, shuffling the data impacts negatively on the generalizability of the model given the limited amount of PL measurements.

\begin{table*}
\renewcommand*{\arraystretch}{1.5}
  \begin{center}
    \caption{{RMSE in PL prediction.}}
    \label{performance_3isto1_train_test}
    {\scriptsize
    \begin{tabular}{|c|c|c|c||c|c|c|}
     \hline
    \multirow{2}{*}{ \textbf{Prediction method}} & \multicolumn{3}{c||}{\textbf{Street-by-street training-testing} } & \multicolumn{3}{c|}{\multirow{1}{*}{\textbf{Links-shuffle-split training-testing} }}\\
     &  \multicolumn{3}{c||}{(mean $\pm$ standard deviation) over 13 streets} & \multicolumn{3}{c|}{(mean $\pm$ standard deviation) over 25 shuffles} \\ 
     \hline
    {\textbf{\emph{3GPP UMi NLOS}}} & \multicolumn{3}{c||}{$\mathbf{18.0 \pm 4.1}$ \textbf{dB}} & \multicolumn{3}{c|}{$\mathbf{18.1 \pm 0.4}$ \textbf{dB}}  \\
    \hline
    {\textbf{\emph{3GPP UMa LOS}}} & \multicolumn{3}{c||}{$\mathbf{10.6 \pm 4.4}$ \textbf{dB}} & \multicolumn{3}{c|}{$\mathbf{11.3 \pm 0.4}$ \textbf{dB}}  \\
    \hline
    {\textbf{\emph{Slope-intercept }}} & \multicolumn{3}{c||}{$\mathbf{6.5 \pm 2.0}$ \textbf{dB}}  & \multicolumn{3}{c|}{$\mathbf{6.1 \pm 0.3}$ \textbf{dB}}  \\
    \hline\hline
    \multirow{3}{*}{\textbf{ML Algorithm}}  & \multirow{3}{*}{\textbf{{Clutter}}} & \multicolumn{2}{c||}{\textbf{{Clutter} + {Building}} } & \multirow{1}{*}{\textbf{{Clutter}}} & \multicolumn{2}{c|}{\textbf{{Clutter} + {Building}} }\\
		 \cline{3-4} \cline{5-7}
		 &   & \textit{Average} over & \textit{Best} out of & \textit{Average} over & \textit{Average} over & \textit{Best} out of\\	
		 &   & $25$ AE runs & $25$ AE runs & $25$ shuffles & $25$ AE runs & $25$ AE runs\\		
    \hline
    \hline
    \textbf{\emph{RF}} & $6.6 \pm 1.8$ dB  & $6.9 \pm 1.9$ dB & $5.8 \pm 1.5$ dB  & $\mathbf{4.1 \pm 0.2}$ \textbf{dB} & $\mathbf{4.3 \pm 0.2}$ \textbf{dB} & $\mathbf{4.0}$ \textbf{dB}\\
    \hline
    \textbf{\emph{SVR}} & $5.4 \pm 1.4$ dB  & $5.8 \pm 1.4$ dB & $4.8 \pm 1.1$ {dB} & $4.4 \pm 0.2$ dB & $4.4 \pm 0.2$ dB & $ {4.0}$  {dB}\\
    \hline
    \textbf{\emph{Lasso}} & $5.7 \pm 1.5$ dB  & $ {5.7 \pm 1.4}$ {dB} & $ {4.8 \pm 1.0}$ {dB} & $5.0 \pm 0.2$ dB & $4.8 \pm 0.2$ dB & $4.4$ dB\\
    \hline
    \textbf{\emph{Elastic-net}} & $\mathbf{5.4 \pm 1.3}$ \textbf{dB}  & $\mathbf{5.7 \pm 1.4}$ \textbf{dB}  & $\mathbf{4.8 \pm 1.1}$ \textbf{dB} & $5.0 \pm 0.2$ dB & $4.8 \pm 0.2$ dB & $4.4$ dB\\
    \hline
    \end{tabular}} \vspace{-0.25cm}
  \end{center}
\end{table*}

Motivated by the large street-by-street variation of measured PL observed from Manhattan measurements~\cite{urban1}, we propose a new way of training-testing referred to as \emph{street-by-street training and testing}, focusing on the extrapolation capabilities. We group the measurement links based on streets where they are collected and formulate $13$ groups, one for each street. We then create $13$ train-test combinations, wherein for each combination, one street is selected for testing and the rest for training. A model is trained and tested $13$ times, using the $13$ train-test combinations independently, producing $13$ RMSE values. We summarize the procedure briefly in Algorithm~\ref{alg:prop}.

\begin{remark}
In Algorithm \ref{alg:prop} the street-by-street variation in prediction is quantified by the standard deviation in RMSE of the $13$ tested streets. It is the metric chosen to measure the generalizability to unseen streets. Thus, the lower the standard deviation, the better is the generalizability.
\end{remark}

\subsection{Model Applicability and Generalizability}
The measurement campaign~\cite{urban1} focused on rooftop-to-same street measurements in urban street canyons. Thus, the trained ML-based PL prediction models proposed in this work are applicable to similar urban street canyons, where BS is placed on the rooftop and UEs move along the same street. The direct path between them would have been in LOS if there were no street clutter/foliage.

The generalizability of the trained model to unseen streets depends on the similarity of street canyons and street clutter density. We have introduced three measures (see Sec.~\ref{sec:contributions}) to enhance the ``generalizability” of trained ML models, namely, street-by-street training and testing to force extrapolation, aggressive feature compression to improve robustness, and simple prediction models to reduce over-fitting. 

The applicability and generalizability of the trained ML models to new urban street canyons can be assessed from three aspects:
\begin{itemize}
\item \emph{Ranges of features} -– The trained ML model is expected to work best/well when the ranges of the features in unseen streets are within or close to the ranges of the training features~\cite{Sculley14}, as summarized in Table~\ref{expert_feat_table}.
\item \emph{Feature importance and sensitivity} -– Not all features are equally important in PL prediction and not all features of unseen streets have a similar range as training data. Thus, it becomes pivotal to analyze feature importance and feature sensitivity~\cite{Breck16}. We defer the analysis to Sec.~\ref{sec:feature_imp} and Appendix~\ref{appendix_2}. 
\item \emph{Distribution of measured PL} –- When sparse PL measurements are available from target streets, we can check if the measured PL distribution is similar to the distribution of training PL samples, as depicted in Fig.~\ref{intro_2}. The trained ML model is expected to work best/well only if the distributions are similar.
\end{itemize} 

\begin{remark}
If the feature ranges of future streets and/or the distribution of measured PL do not align well with existing training datasets, we can employ transfer learning~\cite{Transferred_ML1, Transferred_ML2} to fine-tune the existing ML model with very few measurements, instead of training the new ML model from scratch by leveraging the trained ML model.  
\end{remark}

\section{Performance Evaluation and Analysis}\label{sec:results} 
In this section, we evaluate the performance of the proposed PL prediction models using the \emph{street-by-street training and testing}. The key performance metric is the mean and standard deviation  of the $13$ RMSE values obtained in street-by-street PL prediction. Our benchmarks are the 3GPP UMa LOS prediction model $\left(P = 28.0 + 22\log_{10}d_{3\text{D}} + 20\log_{10}f_c\right)$, 3GPP UMi NLOS prediction model $\left(P = 22.4 + 35.3\log_{10}d_{3\text{D}} + 21.3\log_{10}f_c\right)$, where $f_c{=}28$ denotes carrier frequency {(in GHz)}, as well as the slope-intercept model described in (\ref{conv_SI_model}) where the slope and intercept parameters are obtained using the same training data subsets as used by the ML-based methods. All of the three models only use the 3D Tx-Rx Euclidean distance $d_{3\text{D}}$ as the input feature and their performance are evaluated using the same testing data subsets as used by ML-based methods. 

Simple ray-tracing-based PL prediction, which includes up to $10$ reflections from the wall and the ground but ignores street-clutter, has predicted stronger signal power than in free space~\cite{TAP_17, urban1}, which itself is $10$~dB hotter than the measured data in urban canyons~\cite{urban1}. Several works~\cite{Lee18, Charbonnier20, Leekim19} have focused on (3D) ray-tracing-based prediction for mm-wave in urban street canyons, where the accuracy of ray-tracing is shown to be strongly dependent on modeled wave propagation mechanisms~\cite{Charbonnier20} and environmental feature description~\cite{Leekim19}. Therefore, an in-depth proper investigation of street clutter approximation and modeling is needed to bring ray-tracing prediction up to a reasonable level of accuracy. We leave improvements of ray-tracing-based PL prediction for future work. 

\subsection{PL Prediction Accuracy (RMSE)}
We summarize in Table~\ref{performance_3isto1_train_test} the RMSE in PL prediction of linear and non-linear ML algorithms proposed in Sec.~\ref{ML_model} with street-by-street training-testing methodology (shown in Algorithm~\ref{alg:prop}), where the standard deviation of RMSE represents robustness against street-by-street variation over all $13$ training-testing combinations.  

The 3GPP UMi NLOS and 3GPP UMa LOS channel models have mean RMSE of  $18.0$ and $10.6$ dB, respectively, not suitable for describing street canyon channels with clutters. The slope-intercept model  produces a mean RMSE of $6.5$~dB with a standard deviation of $2.0$~dB  across different testing streets. With the \textit{Clutter} feature set ($\mathcal{F}_{\text{Clutter}}$), both regularized linear Elastic-net model and the non-linear SVR model simultaneously reduce the mean RMSE by about $1.1$~dB and street-by-street standard deviation by about $0.7$~dB, creating a more generalizable model with better PL prediction accuracy. 

\begin{figure}[t!]
\centering\hspace*{-0.4cm}
\includegraphics[scale=0.5]{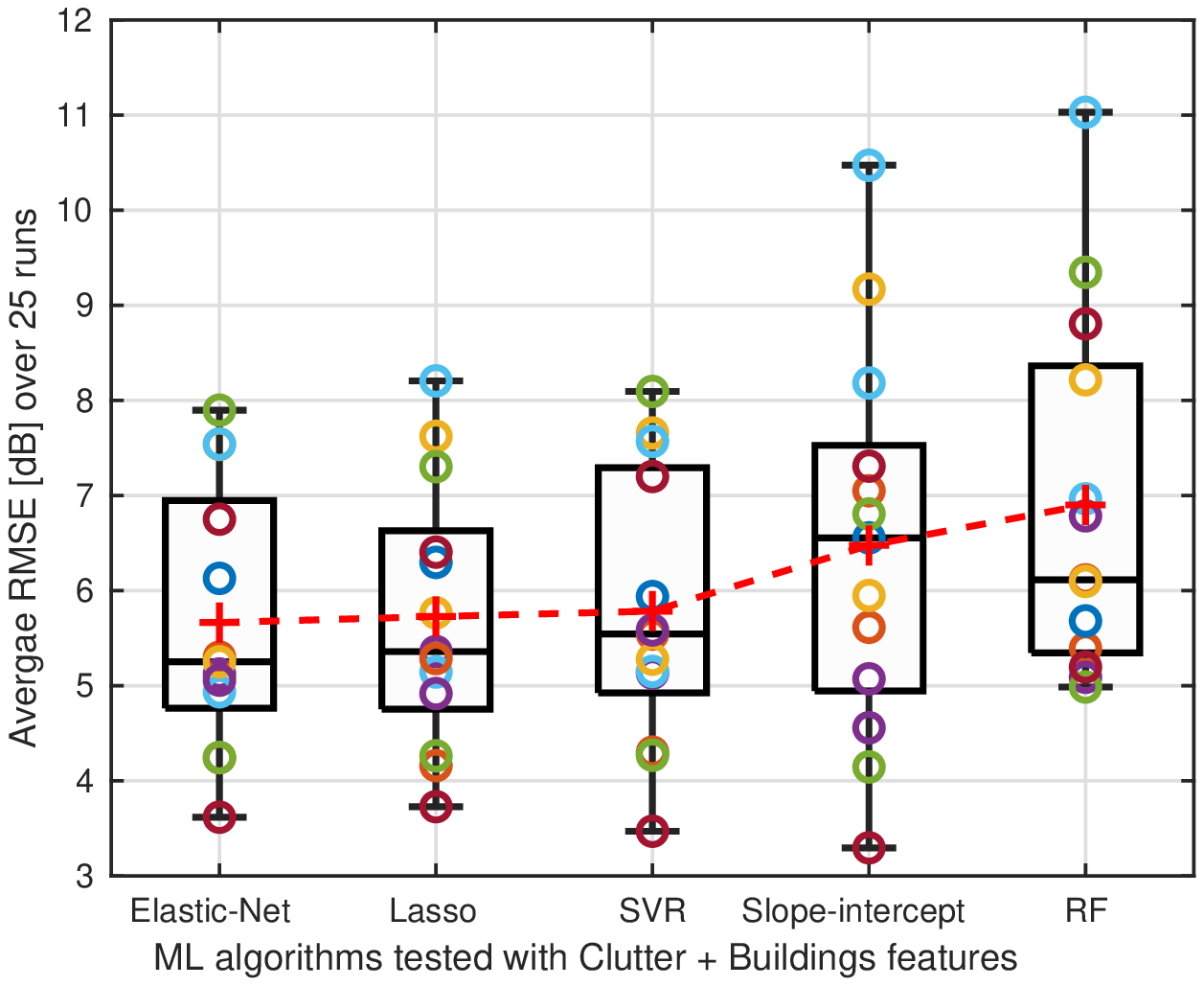}\hspace*{-0.45cm}
\includegraphics[scale=0.6]{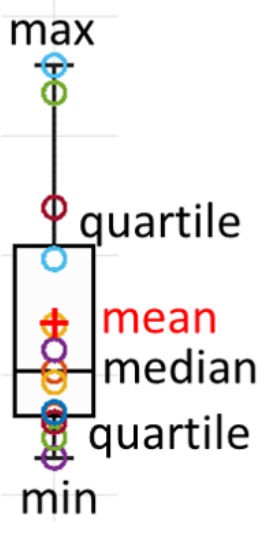} 
\caption{{Street-by-street variation of the average RMSE over 25 runs using \textit{Clutter} + \textit{Building}.}\vspace{-0.35cm}}
\label{dl1_perf}
\end{figure}

The PL prediction performance can be further improved using the $\mathcal{F}_{\text{Clutter\_Building}}$ feature set. Unlike the \textit{Clutter} feature set which is deterministic, the $\mathcal{F}_{\text{Clutter\_Building}}$ feature set contains compressed building features extracted from a CNN-based AE, which is inherently a random process and the resulting performance can change significantly~\cite{Reproduce2018}. We run the AE $25$ times and generate $25$ unique $\mathcal{F}_{\text{Clutter\_Building}}$ feature sets. For each of the $25$ feature sets, we test the ML algorithms using the street-by-street testing, reporting both the best\footnote{Both the mean and standard deviation of the best RMSE converge  within 25 runs as shown in Appendix~\ref{AE_convergence_appendix}.} and the average over all $25$ runs. Street-by-street variation of the average RMSE over all $25$ runs are reported in Fig.~\ref{dl1_perf} using a box plot, where the average RMSE over 25 runs for each testing street is represented as a color-coded $o$ symbol. The median and mean over $13$ streets are given by line inside each box and the red $+$ symbol, respectively, and the edges of the box mark the quartiles, with whiskers extending outside the box indicating the minimum and maximum over all $13$ testing streets.  By searching for a better AE out of $25$ runs for each testing street, the mean RMSE can be further reduced by about $0.6$~dB and street-by-street standard deviation by about $0.2$~dB for both Elastic-net and SVR. The gain of Random Forest-based prediction over slope-intercept model is small, which is likely because it is not good at extrapolation when the statistics of the training and testing sets differ~\cite{RF_limit}. 

To verify and compare the capability of interpolation of various prediction models, we also run the classical links-shuffle-split training and testing approach detailed in Algorithm~\ref{alg:conv} with $iterations{=}25$, i.e., randomly shuffle-and-split $25$ times.  The results are also shown in Table~\ref{performance_3isto1_train_test}, where the $0.2$ dB standard deviation in RMSE for \textit{Clutter} + \textit{Building} comes from both the inherent randomness of shuffling-splitting the dataset and AE compression out of $25$ independent runs. The best result out of 25 runs for all the proposed ML-based PL prediction achieved over $1.7$ dB reduction in mean RMSE  as compared to the slope-intercept prediction, and over 0.4 dB reduction compared to the street-by-street testing. However, caution has to be taken to differentiate extrapolation and interpolation for site-specific PL prediction given limited measurement datasets.

\subsection{Robustness Against Street-by-Street Variation and Distance}
\begin{figure}[t!]
\centering\hspace*{-0.35cm}
\includegraphics[scale=0.4]{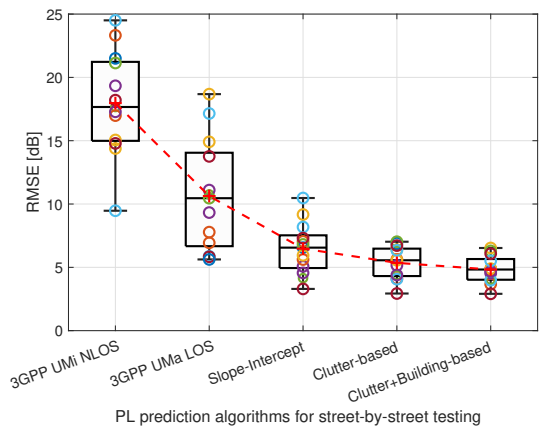}
\caption{Street-by-street variation.\vspace{-0.35cm}}\label{street_by_street_summary}
\end{figure}

We evaluate the robustness of PL prediction against street-by-street variation in Fig.~\ref{street_by_street_summary}  using Elastic-net-based prediction for both the \textit{Clutter} feature set and the best \textit{Clutter + Building} $(\mathcal{F}_{\text{Clutter\_Building}})$ feature set out of the $25$ options. Compared to the two 3GPP models and the classical slope-intercept model, the two ML-based PL prediction models reduce both the mean RMSE as well as the street-by-street variability. This clearly demonstrates the importance of incorporating street-specific features such as street clutter and building into PL prediction models and the robustness of linear ML-based models in extrapolation to unseen streets.

\begin{figure}[t!]
\centering\hspace*{-0.35cm}
\includegraphics[scale=0.42]{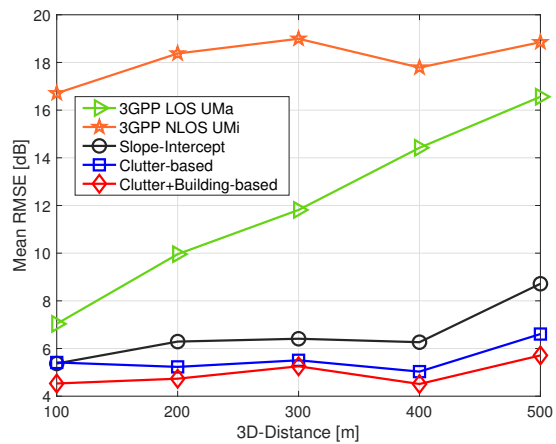}
\caption{ Mean RMSE versus distance. \vspace{-0.35cm}}\label{dist_perf}
\end{figure}

\begin{figure*}[t]
\centering\hspace*{-0.15cm}
\begin{subfigure}[t]{0.31\textwidth}
\centering
\includegraphics[scale=0.38]{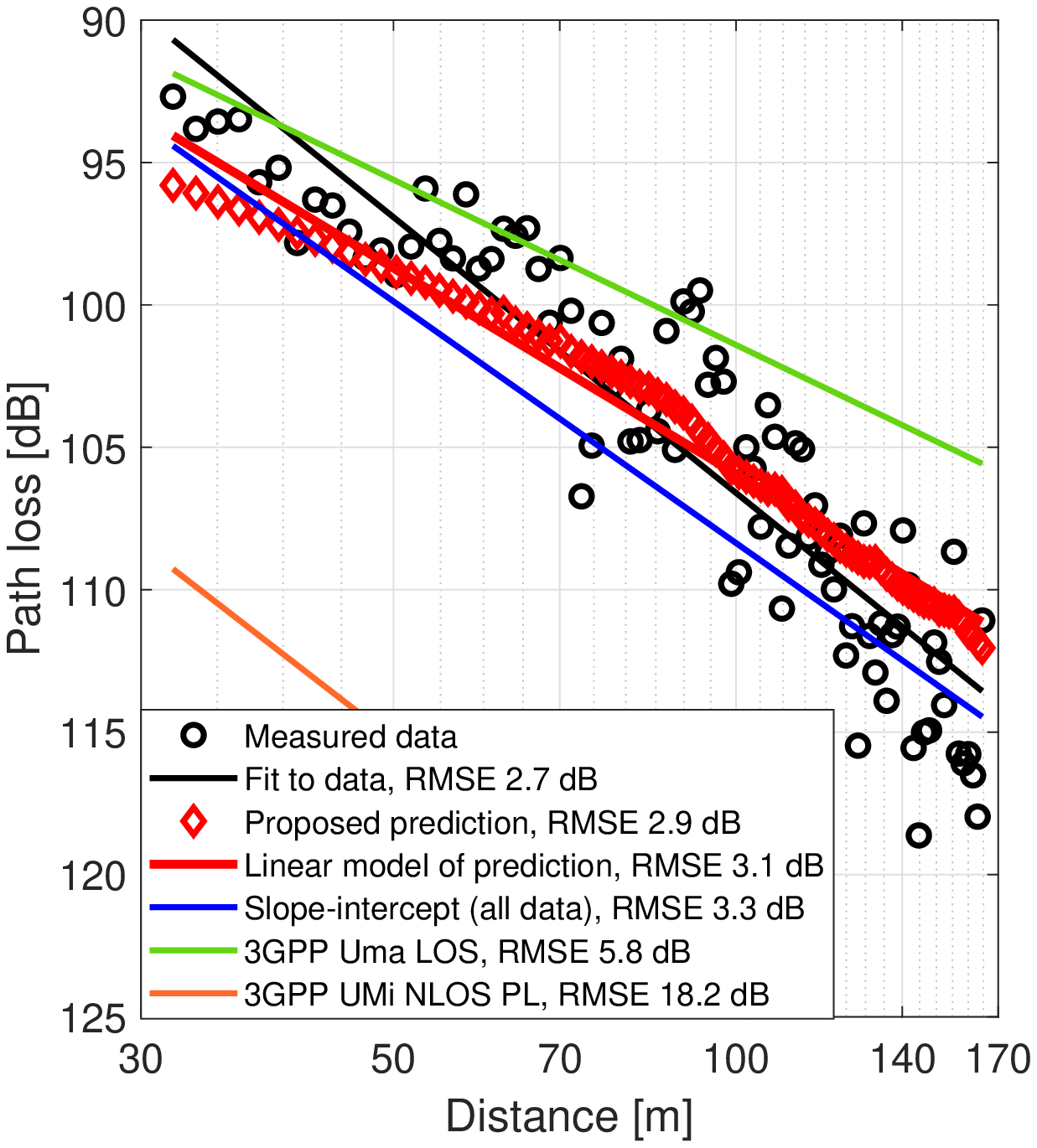}
\caption{{Street-by-street - Best case (Street 1).}}
\label{sbs_1}
\end{subfigure}\hspace*{0.35cm}
\begin{subfigure}[t]{0.31\textwidth}
\centering
\includegraphics[scale=0.38]{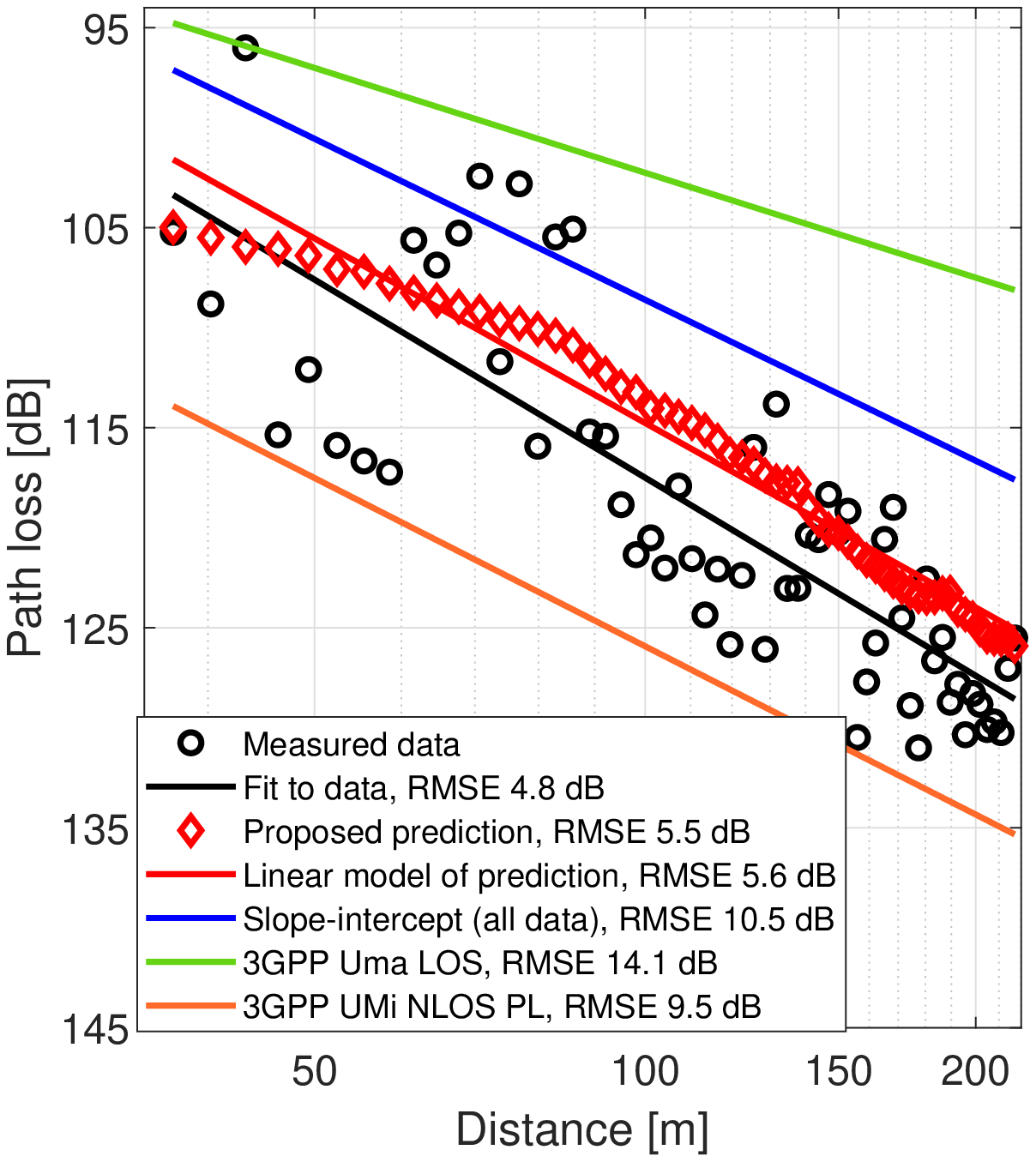}
\caption{{Street-by-street - Worst case (Street 13).}}
\label{sbs_2}
\end{subfigure}\hspace*{0.35cm}
\begin{subfigure}[t]{0.31\textwidth}
\centering
\includegraphics[scale=0.38]{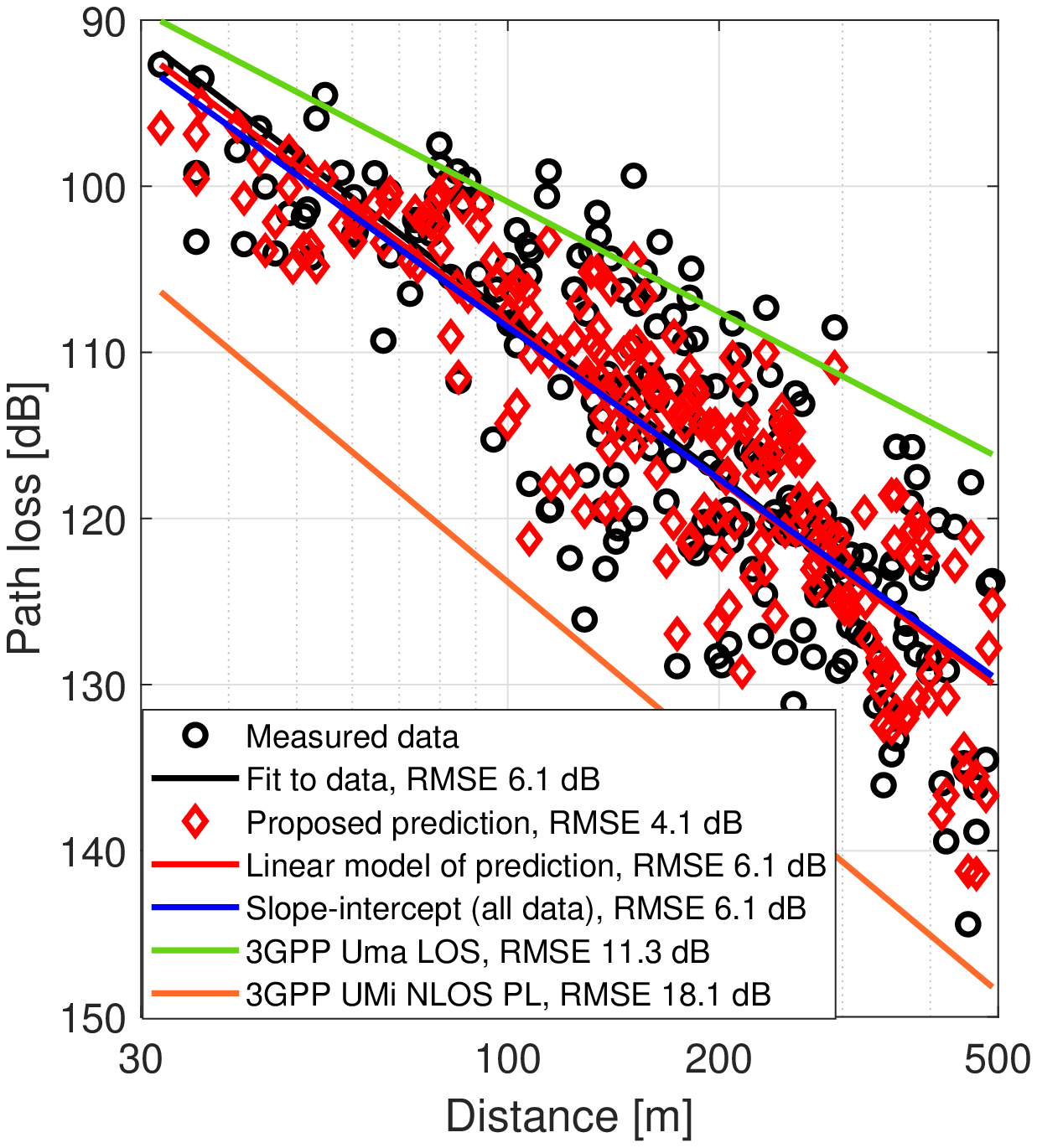}
\caption{{Link-shuffle-split with {80\%-20\% split}.}}
\label{lss}
\end{subfigure}
\caption{Measured PL data versus prediction PL data. For the street-by-street testing (in Fig. a, b) we employ Elastic-Net on \textit{Clutter + Building} features. For the link-shuffle-split testing (in Fig. c) we employ random forest (RF) on \textit{Clutter} features.\vspace{-0.35cm}}
\label{sbs_lss}
\end{figure*}

The mean RMSE in PL prediction as a function of the 3D distance is shown in Fig.~\ref{dist_perf}. All the links are grouped into $100$~m intervals and  within each interval  we calculate the mean RMSE for street-by-street testing. Herein, X-axis denotes the ending position of an interval (e.g., $200$~m denotes the interval spanning from $100$ to $200$~m). Our proposed models outperform the slope-intercept model for all the distance ranges. The improvement in mean RMSE is about $1.2$ to $2.7$~dB using \textit{Clutter + Building} feature set compared to the classical slope-intercept model. 

\subsection{PL Prediction versus Measurement}
We compare in Fig.~\ref{sbs_lss} the measured PL values with the predicted PL values. We analyze the street-by-street testing using Elastic-net prediction (regularized linear regression) with \emph{Clutter} + \emph{Building} features for the best-case (smallest  RMSE, \textit{Street 1}) and the worst-case (largest RMSE, \textit{Street 13}) in Fig.~\ref{sbs_1} and Fig.~\ref{sbs_2}, respectively. 
The linear model derived from the Elastic-net prediction (solid red line) is very close to the slope-intercept fit to measured data (solid black line), with marginally increased RMSE ranging between $0.2-0.7$~dB. This indicates that our proposed Elastic-net is highly generalizable and effective in capturing street-by-street variation. While the trained model is good at capturing the overall PL trend per street, its capability of tracking link-by-link variation is modest, as shown by the marginal improvement ($0.1-0.2$~dB in RMSE) over  the linear models derived from the corresponding prediction. In Fig.~\ref{lss} we also analyze the links-shuffle-split testing over 20\% of measured data from all 13 streets using the trained Random Forest with \emph{Clutter} features. This shows that our proposed Random Forest (non-linear regression) model is highly effective in generalizing to nearby links.

\subsection{Feature Importance for the Point cloud-based Street Clutter Features}\label{sec:feature_imp}
\begin{figure}
\centering
\includegraphics[scale=0.485]{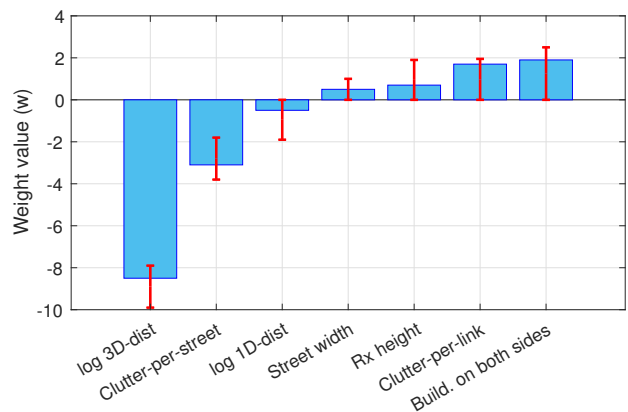} 
\caption{Average feature importance for \textit{Clutter} feature set using Lasso regression in street-by-street testing, where the error bars indicate the minimum-maximum range over 13 streets.\vspace{-0.35cm}}\label{feat_imp}
\end{figure} 

The Lasso regression imposes an $l_1$-norm penalty that minimizes the weights of least relevant features to improve the accuracy, and therefore the relative magnitude of the resulting weights can be interpreted as the feature importance in PL prediction. We perform Lasso regression on the \textit{Clutter} feature set with street-by-street training-testing methodology and present the obtained weights in Fig.~\ref{feat_imp}, where each bar represents the mean value of the weights corresponding to each feature. The error bar indicates the minimum and maximum weight value obtained for that specific feature when tested separately for $13$ streets. The amplitude of each weight indicates the importance of that feature, and the opposite signs (in amplitude) of the weights help balance the \textit{Clutter} features in PL prediction. As expected, the \emph{3D distance} has the highest weight, followed by the \emph{Clutter-per-street}, which quantifies how cluttered each street is based on the normalized point cloud densities of the whole street. The \emph{`Build. on both sides'} {binary} indicator  and the \emph{Clutter-per-link} which quantifies accumulated clutter density along the direct path also have notable importance. This aligns with the understanding that reflection from buildings on both sides of the canyon increases signal strength, and that clutter intruded into the direct path has adversarial effect on propagation. 

  Feature importance can also be observed by comparing the change of prediction accuracy when each \textit{Clutter} feature is excluded individually from prediction, as shown in Appendix~\ref{appendix_2} using Elastic-net regression, where same top four most important features are identified (in descending order): \emph{Clutter-per-street}, \emph{3D distance},  \emph{Clutter-per-link}, and \emph{building on both sides}. 
	
	By only using the top four most influential features, RMSE of $5.5\pm 1.1$~dB can be achieved using Elastic-net regression. For almost all the ML-based prediction results using the reduced \textit{Clutter} feature set,  both the mean RMSE and standard deviations are within $0.2$ dB from the results shown in Table~\ref{performance_3isto1_train_test} obtained using all the seven \textit{Clutter} features. See Appendix~\ref{appendix_2} for details.

\subsection{Computational Complexity}
\begin{table}
\centering
\caption{Training and testing complexity of PL prediction algorithms.}
    \label{complexity_table}
 {\footnotesize
    \begin{tabular}{|c|c|c|}
     \hline
     \textbf{Algorithm} & \textbf{Training Complexity} & \textbf{Testing Complexity} \\
     \hline
     \hline
     \emph{3GPP} & $-$ & \multirow{2}{*}{$\mathcal{O}\left(1\right)$}\\
     \cline{1-2}
     \emph{Slope-intercept} & $\mathcal{O}\left(n\right)$ & \\
     \hline
     \emph{Lasso} & $\mathcal{O}\left(Tp^2n+p^3\right)$ & \multirow{2}{*}{$\mathcal{O}\left(p\right)$} \\
     \cline{1-2}
     \emph{Elastic-Net} & $\mathcal{O}\left(Tp^2n+p^3\right)$ & \\
     \hline
     \emph{Random forest} & $\mathcal{O}\left(n\log(n)pn_{t}n_{d}\right)$ & $\mathcal{O}\left(n_{t}n_{d}\right)$ \\
     \hline
     \emph{Support Vector} & $\mathcal{O}\left(n^2p + n^3\right)$ & $\mathcal{O}\left(n_{sv}p\right)$ \\
     \hline
    \end{tabular}} \vspace*{-0.35cm}
    \end{table}
    
Let us denote the number of training samples by $n$, number of features by $p$, number of trees by $n_t$ and depth of the tree by $n_d$, and number of support vectors by $n_{sv}$,  the training and testing complexity can be summarized as in Table IV where $T$ is the number of outer iterations used in the coordinate descent solver. Given the low dimensionality of the features (7 in \textit{Clutter} and 12 in \textit{Building}) used in our PL prediction models, the PL prediction is very fast. The prediction time per link is less than 3 $\mu s$ for Lasso and Elastic-net and less than 40 $\mu s$ for RF and SVR when 
implemented using scikit-learn~\cite{scikit} running with Intel Core i7--6700 CPU with 64 GB RAM and Ubuntu 18.04.4 LTS OS.

\section{Conclusions}\label{sec:conclusion} 
We have proposed a ML-based PL prediction model for urban street canyon using the $28$ GHz measurement data collected from Manhattan. The feature set contains street clutter obtained from LiDAR point cloud and  buildings from 3D mesh-grid. The PL dataset has 1028 PL measurement links from 13 streets. Although the PL dataset is massive for the classical slope-intercept PL modeling, it is small for ML-based approaches when compared to massive point cloud and 3D building feature sets and parameters in AE and learning algorithms. To mitigate the risk of overfitting, we defined seven expert features with physical meaning from the point cloud. We also compressed 3D building features to a length-12 vector for each link using CNN-based AE. Instead of interpolation to nearby links, we focused on the extrapolation by introducing a street-by-street training and testing approach. Using linear ML algorithms for PL prediction, we achieved RMSE of $4.8 \pm 1.1$ dB compared to $10.6 \pm 4.4$ dB and $6.5 \pm 2.0$ dB for 3GPP LOS and slope-intercept prediction, respectively, which demonstrates the superior capability of our model in extrapolation.

Intuitive interpretation of feature importance was obtained using Lasso regression-based analysis and feature-exclusion analysis. By only using the top four most influential features, namely, distance, street clutter density (\emph{Clutter-per-street} and \emph{Clutter-per-link}), and street canyon indication (\textit{building on both sides}), RMSE of $5.5\pm 1.1$~dB can be achieved using Elastic-net regression.
  
 Among the four learning algorithms used in this paper, the non-linear Random Forest regression has achieved the worst prediction performance under street-by-street testing but is the best under links-shuffle-split testing. This may be attributed to the reduced similarity in statistics between training and testing sets in street-by-street testing compared to the links-shuffle-split testing, and the lack of extrapolation capabilities of Random Forest regressions. 
Regularized linear Elastic-net regression has the best performance, which is in line with the intuition that regularization on linear algorithms is more robust against overfitting with limited training dataset. The non-linear Support Vector Regression with radial bias function kernel performs well (second best) for both links-shuffle-split testing and street-by-street testing, at the cost of higher complexity.

\appendices
\section{CNN-based Autoencoder design to Compress Building Features}\label{appenda} 
We design CNN-based AE for feature extraction from the building dataset, as described in Fig. \ref{dl_arch}, where an encoder $e(\cdot)$ compresses the input 2D building collapse $\mathbf{I}\in\mathbb{R}^{(500, 40)}$ to an representation $\textbf{X}\in\mathbb{R}^{(12)}$, which is then fed to the decoder $d(\cdot)$ to reconstruct original input 2D building collapse $\mathbf{I}$. We briefly describe below the network layers shown in Fig.~\ref{dl_arch_0}.
\begin{enumerate}
\item \textit{1D-convolutional layer (Conv 1D)} - It employs various kernels to convolve the 2D-image, preserving the spatial characteristics of the input image while extracting relevant features.
\item \textit{1D-max-pooling layer (Max-Pooling 1D)} - Pooling is a sample-based discretization process utilized to downsample the input image by making assumptions in the binned sub-region. In max-pooling, we take the maximum value in the binned sub-region. 
\item \textit{1D-upsampling layer (Up-Sampling 1D)} -  Upsampling layer has no weights, which helps increase the input dimensions when followed by a convolutional layer.
\item \textit{Fully connected layer (Dense)} - The non-linear processing is performed via dense layers wherein each neuron is fully connected to all the neurons in the previous layer. 
\end{enumerate}
Please note that 1D in the above layers means we have kernels (in Conv layer) and factor (for upsampling and downsampling)~\cite{DLbook} in only one-dimension. 

\vspace*{-0.2cm}
\subsection{Designing the Encoder}
The input to our network is 2D collapse of buildings given by $\mathbf{I}\in\mathbb{R}^{(500, 40)}$, which is normalized between $0$ and $1$. Let $e(\mathbf{I}|\Theta_e)$ be the mapping from the input buildings to compressed representation when the parametric transformation of encoder is given by $\Theta_e$, which denotes the weight, filters, and bias terms. Thus the encoder can be denoted as
\begin{align}
\textbf{X} = e(\mathbf{I}|\Theta_e) = e_{L_e}\left(...(e_2\left(e_1\left(\textbf{I}|\Theta_{e_1}\right)|\Theta_{e_2}\right)...|\Theta_{e_{L_e}}\right)
\end{align}
where $L_e$ denotes the number of layers in the encoder. 

For the convolutional layers, the operation of the $l^\mathrm{th}$ layer can be represented as follows
\begin{align}
\textbf{X}_l = e_l(\mathbf{I}_l|\Theta_{e_l}) = h\left(\mathbf{W}_l \otimes \mathbf{I}_l + b_l\right), \quad \Theta_{e_l} = [\mathbf{W}_l, b_l]
\end{align}
where $\otimes$ indicates the convolutional process, $\mathbf{W}_l$ represents the {1D} kernels used for feature extraction, $b_l$ denotes the bias vector, $h(\cdot)$ is the activation function, $\mathbf{I}_l{=}\textbf{X}_{l-1}$ comes from layer concatenation and $\mathbf{I}_1$  equals the {2D} building matrix $\mathbf{I}$.

 We apply several \emph{Max-Pooling 1D} layers in between for improving the region covered by the following receptive fields. Moreover, as shown in Fig. \ref{dl_arch_1}, we introduce grouped convolutions in the encoder, wherein we take the output of the first \emph{Max-Pooling 1D} and make two branches of it, with separate \textit{Conv 1D} and \emph{Max-Pooling 1D} layers, and then add the output of both branches (inspired by AlexNet \cite{DLbook}). The convolutional layer's output is then flattened to $\mathbf{K}$ and used as input of several stacked dense layers, where the first dense layer in the encoder can be given by
\begin{align}
\textbf{X}_l = e_l(\mathbf{I}_l|\Theta_{e_l}) = h\left(\mathbf{W}_l\mathbf{K} + b_l\right), \quad \Theta_{e_l} = [\mathbf{W}_l, b_l]
\end{align}

\subsection{Designing the Decoder}
The input to our decoder is the output of the encoder given by $\textbf{X}\in\mathbb{R}^{(12)}$. Let $d(\mathbf{X}|\Theta_d)$ be the mapping from the compressed representation to input buildings when the parametric transformation of decoder is given by $\Theta_d$, which denotes its weight, filters, and bias terms. The decoder can be represented as follows
\begin{align}
\textbf{Y} = d(\mathbf{X}|\Theta_d) = d_{L_d}\left(...(d_2\left(d_1\left(\textbf{X}|\Theta_{d_1}\right)|\Theta_{d_2}\right)|\Theta_{d_{L_d}}\right)
\end{align}
where $L_d$ denotes the number of layers in the decoder. It performs a reverse operation of encoder here to generate the output $\textbf{Y}\in\mathbb{R}^{(500,40)}$ of the same size as the input $\textbf{I}$. 

As shown in Fig. \ref{dl_arch_0}, we use the \emph{Tanh} activation function $tanh(x) = \frac{e^{x}-e^{-x}}{e^{x}+e^{-x}}$ for all the layers because \emph{Tanh} activation function performed the best compared to the other non--linear activation functions, except for the last layer, where we used the \emph{ReLU} activation function $ReLU(x) = \max(0, x)$ to ensure a positive real-value output. Moreover, $a\times b$ on each \emph{Conv 1D} layer indicates the filters and kernel size. The value on each \emph{Max-Pooling 1D} and \emph{Up-sampling 1D} denotes the factor by which downsampling and upsampling are performed on the first dimension. Also, value on each \emph{Dense} layer indicates the number of neurons considered in that layer. The symbol $+$ indicates the addition of the outputs of two previous layers. 

\subsection{Designing the Loss Function}
We use log-cosh loss, which is the logarithm of the prediction error's hyperbolic cosine. Also, we have $\mathbf{I}$ as the input to the encoder in the AE as well as the ground truth to be predicted from the decoder and $\mathbf{Y}$ as the predicted output of the AE. Thus the difference between the input and output of the AE can be given by $\theta_{i,j}=\mathbf{Y}_{i,j}-\mathbf{I}_{i,j}, \,\forall\,(i,j)$, where $i = \{1, ..., 500\}$ and $j = \{1, ..., 40\}$ denotes the length and width of the streets (with building facades). We choose log-cosh loss to help stabilize the training performance with fewer epochs (iterations) because the outliers minimally impact the log-cosh loss compared to the MSE loss~\cite{DLbook}. Also, since we have appended zeros in the input $\mathbf{I}$,  there have many zeros appended for shorter distances, which makes it difficult for the AE network to learn non-zeros values in closer distance ranges. Thus, we introduce a matrix $\hat{\mathbf{Y}}\in\mathbb{R}^{(500, 40)}$, where for the $n^\mathrm{th}$ training sample, given by
\begin{align}
\hat{\mathbf{Y}}^n_{i,j} = \begin{cases}
0, \qquad\quad\;\;\,\;\;\text{if}\quad\mathbf{I}^n_{i,j} = 0 \\
\mathbf{Y}^n_{i,j}, \qquad\quad\, \text{otherwise}
\end{cases}, \quad \forall\;(i,j)
\end{align}
 Then the combined loss function for the $n^\mathrm{th}$ training sample can be given by 
\begin{align}
\mathcal{L} &= \mu\left(\log\left(\cosh\left(\hat{\mathbf{Y}}^n-\mathbf{I}^n\right)\right)\right) + \nonumber\\
&\qquad\qquad\qquad 0.1\times\mu\left(\log\left(\cosh\left(\mathbf{Y}^n-\mathbf{I}^n\right)\right)\right)\label{AEloss}
\end{align}
where $\mu(\cdot)$ is the mean. The loss function has two parts, the first part focuses on the reconstruction error of the non-zero values. The second part focuses on the reconstruction error of all the values, whereas the weight $0.1$ helps us in reducing the impact of the appended zeros.

\begin{figure}[t]
\centering\hspace*{-0.35cm}
\begin{subfigure}[t]{0.24\textwidth}
\centering
\includegraphics[scale=0.35]{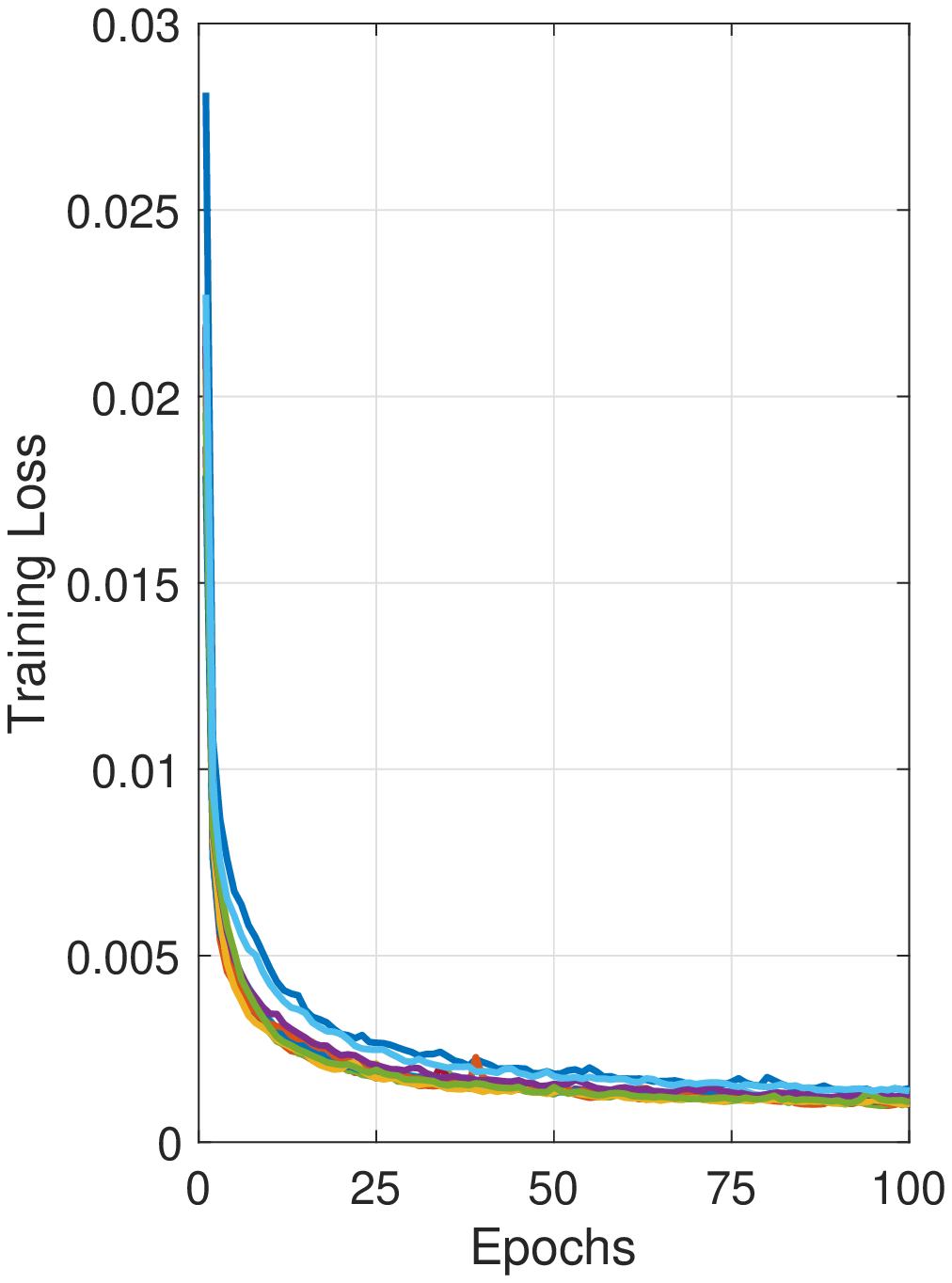}
\caption{Training loss.}
\label{AE_1}
\end{subfigure}
\begin{subfigure}[t]{0.24\textwidth}
\centering
\includegraphics[scale=0.35]{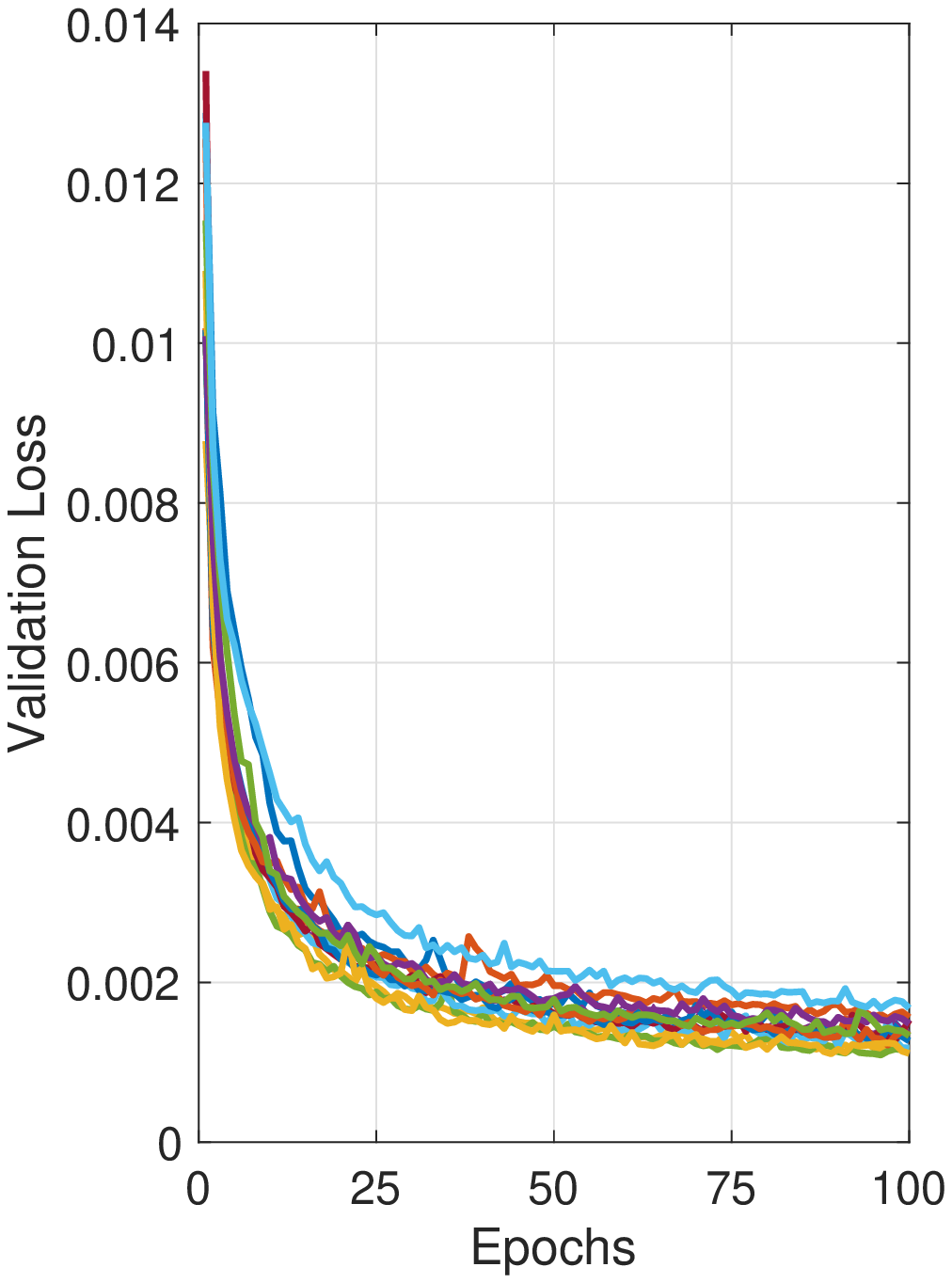}
\caption{Validation loss.}
\label{AE_2}
\end{subfigure}
\caption{Convergence of {CNN AE} for 13 testing streets.\vspace{-0.45cm}}
\label{AE_converg}
\end{figure} 

\begin{table*}[t]
\renewcommand*{\arraystretch}{1.5}
  \begin{center}
    \caption{{RMSE in PL prediction using the 4 most important Clutter features (\textit{Clutter$_4$}).}}
    \label{performance_top_4_PCEF}
    {\scriptsize
    \begin{tabular}{|c|c|c|c||c|c|c|}
    \hline
     & \multicolumn{3}{c||}{\textbf{Street-by-street training-testing} } & \multicolumn{3}{c|}{\multirow{1}{*}{\textbf{Links-shuffle-split training-testing} }}\\
     \hline
     \hline
    \multirow{3}{*}{\textbf{ML Algorithm}}  & \multirow{3}{*}{\textbf{{Clutter$_4$}}} & \multicolumn{2}{c||}{\textbf{{Clutter$_4$} + {Building}} } & \multirow{1}{*}{\textbf{{Clutter$_4$}}} & \multicolumn{2}{c|}{\textbf{{Clutter$_4$} + {Building}} }\\
		 \cline{3-4} \cline{5-7}
		 &   & \textit{Average} over & \textit{Best} out of & \textit{Average} over & \textit{Average} over & \textit{Best} out of\\	
		 &   & $25$ AE runs & $25$ AE runs & $25$ shuffles & $25$ AE runs & $25$ AE runs\\		
    \hline
    \hline
    \textbf{\emph{RF}} & $6.2 \pm 1.4$ dB  & $6.9 \pm 1.9$ dB & $5.8 \pm 1.5$ dB & $\mathbf{4.1 \pm 0.2}$ \textbf{dB} & $\mathbf{4.3 \pm 0.2}$ \textbf{dB} & $\mathbf{3.9}$ \textbf{dB}\\
    \hline
    \textbf{\emph{SVR}} & $5.6 \pm 1.2$ dB  & $5.7 \pm 1.3$ dB & $4.8 \pm 0.9$ dB & $4.3 \pm 0.2$ dB & $ {4.3 \pm 0.2}$  {dB} & $4.0$ dB\\
    \hline
    \textbf{\emph{Lasso}} & $5.7 \pm 1.5$ dB  & $5.7 \pm 1.3$ dB & $4.8 \pm 1.0$ dB & $5.1 \pm 0.2$ dB & $4.8 \pm 0.2$ dB & $4.5$ dB\\
    \hline
    \textbf{\emph{Elastic-net}} & $\mathbf{5.5 \pm 1.1}$ \textbf{dB}  & $\mathbf{5.5 \pm 1.2}$ \textbf{dB} & $\mathbf{4.7 \pm 1.0}$ \textbf{dB} & $5.1 \pm 0.2$ dB & $4.8 \pm 0.2$ dB & $4.5$ dB\\
    \hline
    \end{tabular}} \vspace*{-0.5cm}
  \end{center}
\end{table*}

\section{{Loss Convergence and Reproducibility of the Proposed CNN-based AE}}\label{AE_convergence_appendix}
We utilize the designed AE to extract the compressed feature vector $\mathbf{X} \in \mathbb{R}^{(12)}$ from the $2$d collapse $\mathbf{I} \in \mathbb{R}^{(500, 40)}$ of the 3D building dataset. Then we train the AE in an end-to-end manner by minimizing the designed loss in  \eqref{AEloss}. The convergence of the training and validation losses for $13$ models created for street-by-street training and testing are shown in Fig.~\ref{AE_converg}, wherein the loss converges within $50$ epochs (iterations utilized by Adam optimizer to converge).

\begin{figure}
\centering
\begin{subfigure}[t]{0.24\textwidth}
\centering
\includegraphics[scale=0.325]{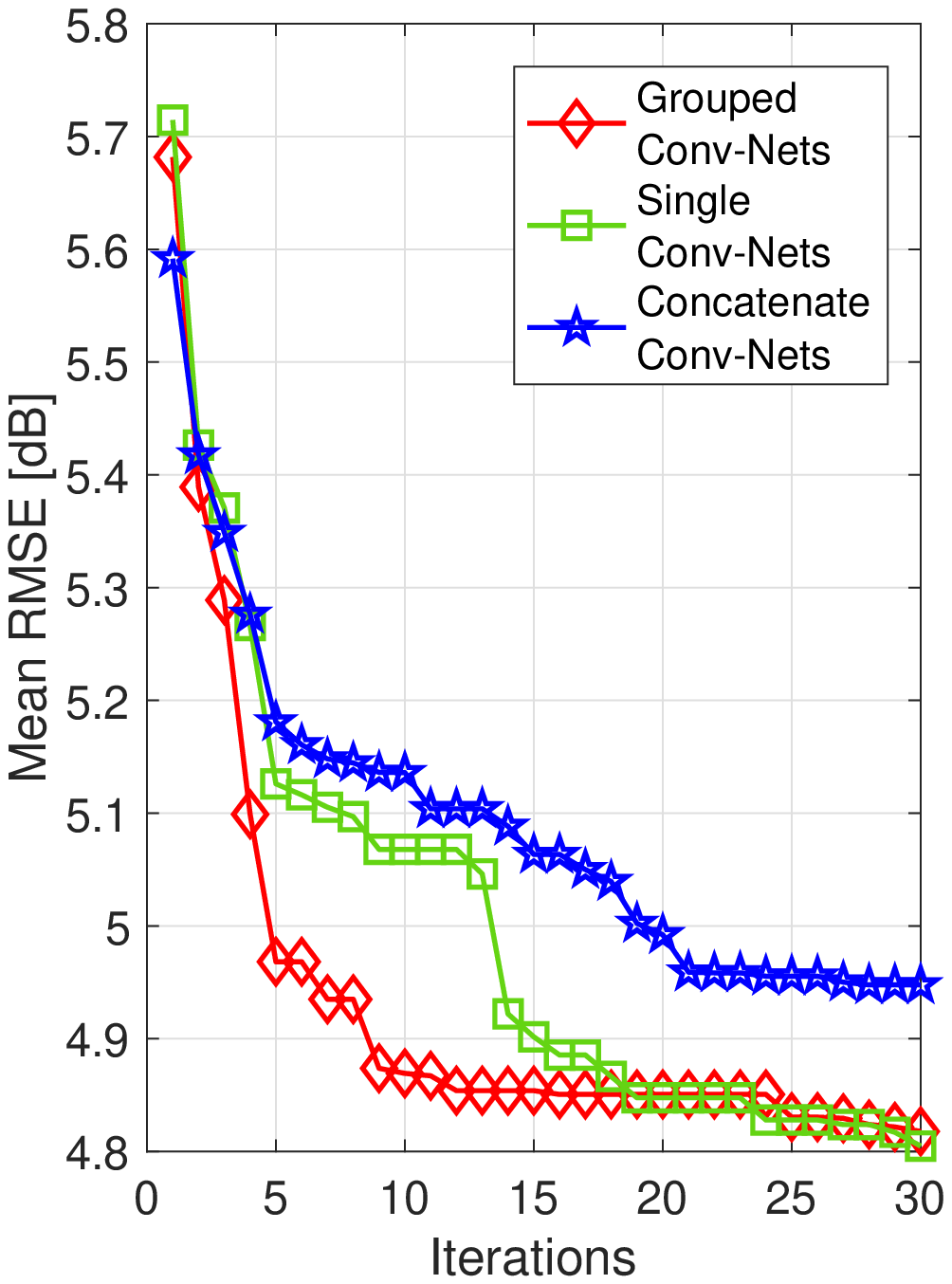}
\caption{$\mu$(RMSE) versus runs.}
\label{review_comm_perf_6}
\end{subfigure}
\begin{subfigure}[t]{0.24\textwidth}
\centering
\includegraphics[scale=0.325]{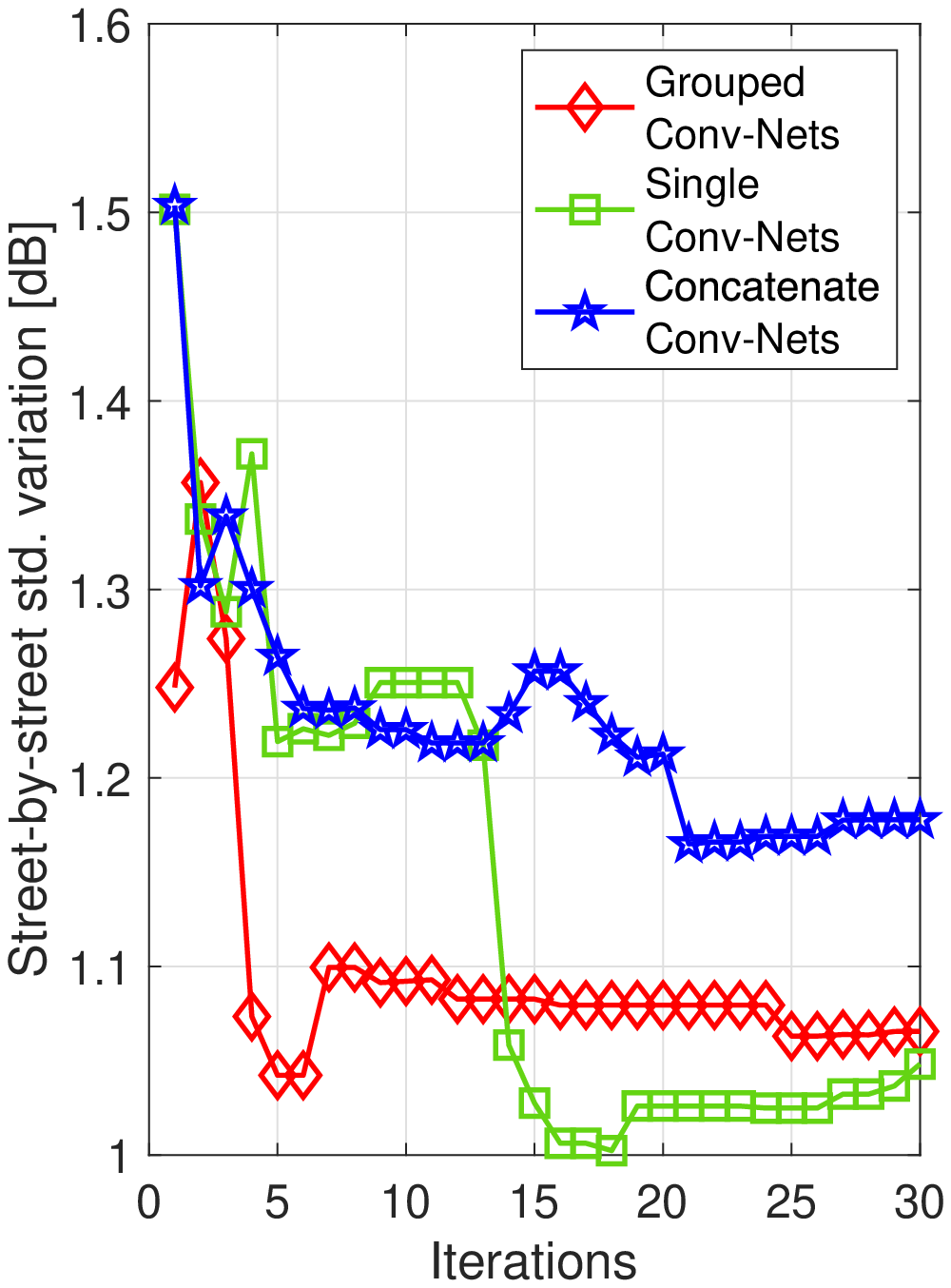}
\caption{$\sigma$(RMSE) versus runs.}
\label{review_comm_perf_7}
\end{subfigure}
\caption{Evaluation of grouped CNNs and iterations needed for reproducibility.\vspace{-0.45cm}}\label{review_comm_perf_8} 
\end{figure}

In Fig. \ref{review_comm_perf_8}, we evaluate the compressed features $(\mathbf{X})$ by making three types of AEs, wherein \emph{only} the grouped CNNs in the encoder as proposed in Fig. \ref{dl_arch_1} is replaced by: (1) Grouped Conv-Nets -- as proposed, (2) Single Conv-Net -- remove \textit{Conv-Net-2} from the encoder, and (3) Concatenate Conv-Nets -- concatenate \textit{Conv-Net-1} and \textit{Conv-Net-2} serially, to obtain their respective feature set $\mathcal{F}_{\text{Clutter\_Building}}$. Further, we show the best RMSE performance achieved by the {Elastic-net} regression to predict the PL, with a varying number of iterations. Grouped CNNs perform the best with smoother convergence compared to others. Furthermore, Fig. \ref{review_comm_perf_8} also shows that with $25$ iterations we can achieve the \emph{reproducibility} for the best PL prediction RMSE performance.

\section{Analyzing the importance of designed {Clutter} in PL Predictions}\label{appendix_2}
In Fig.~\ref{feat_imp}, we utilized $l_1$-norm based Lasso regression to determine the importance of the individual \textit{Clutter} feature if all seven \textit{Clutter} features are provided for the PL prediction. The importance of features, as quantified by their Lasso weights, in the descending order is given as: \textit{log 3D}-distance, \textit{{Clutter-per-street}}, \textit{building on both sides}, \textit{{Clutter-per-link}}, \textit{Rx height}, \textit{street width}, and \textit{log 1D}-distance.
  
Feature importance can also be observed by comparing the change of prediction accuracy when each \textit{Clutter} feature is excluded individually from prediction. This approach works for all ML-based prediction algorithms and thus provides an alternative way of assessing feature importance. 

\begin{figure}[t!]
\centering
\includegraphics[scale=0.41]{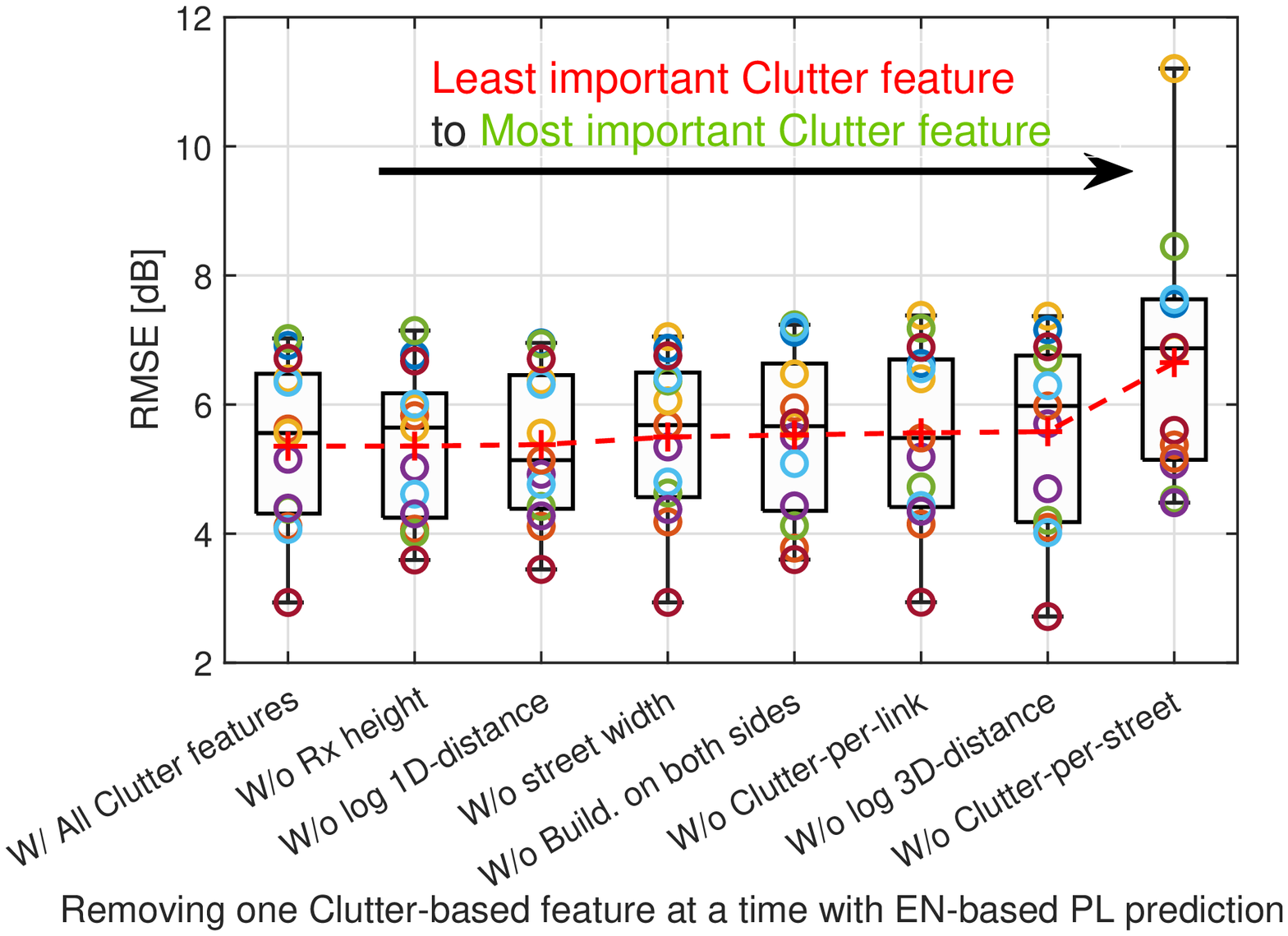}
\caption{Importance of each feature in Elastic-net-based PL prediction model.\vspace{-0.45cm}}\label{En_imp}
\end{figure}

In Fig.~\ref{En_imp}, we remove one feature at a time from the $7$ \textit{Clutter} features and determine the RMSE in Elastic-net based PL prediction using street-by-street testing in Algorithm~\ref{alg:prop}.  Removing the \textit{{Clutter-per-street}} feature has the strongest consequence, degrading the mean RMSE by as much as $1.3$~dB. Thus, based on the degradation of mean RMSE, the importance of the features\footnote{The potential correlation among multiple features may underestimate the importance of a feature if it has high correlation with others.} in descending order can be given: \textit{Clutter-per-street}, \textit{log 3D}-distance, \textit{Clutter-per-link}, \textit{building on both sides}, \textit{street width}, \textit{log 1D}-distance, and \textit{Rx height}.

Based on observations from Fig.~\ref{feat_imp} and Fig.~\ref{En_imp}, we conclude that the most important features are the \textit{log-3D}-distance, street-clutter information given by \textit{Clutter-per-street} and \textit{Clutter-per-link}, and canyon status (\textit{buildings on both sides}), referred as \textit{Clutter}$_4$. We analyze the RMSE in PL prediction using \textit{Clutter}$_4$ for both street-by-street and shuffle-split testing, shown in Table~\ref{performance_top_4_PCEF}. By only using the top four most influential features, RMSE of $5.5\pm 1.1$~dB can be achieved. When using \textit{Clutter}$_4$, both the mean RMSE and standard deviations of almost all the ML-based predictions are within $0.2$ dB from the results where all 7 clutter features are used.
 

\end{document}